\documentclass[12pt,preprint]{aastex}
\shorttitle{}
\shortauthors{Nesvorn\'y et al.}
\begin{document}
\title{Neptune's Orbital Migration Was Grainy, Not Smooth}
\author{David Nesvorn\'y$^1$, David Vokrouhlick\'y$^2$}
\affil{(1) Department of Space Studies, Southwest Research Institute, 1050 Walnut St., \\Suite 300, 
Boulder, CO 80302, USA} 
\affil{(2) Institute of Astronomy, Charles University, V Hole\v{s}ovi\v{c}k\'ach 2, \\
180 00 Prague 8, Czech Republic}

\begin{abstract}
The Kuiper belt is a population of icy bodies beyond the orbit of Neptune. The complex orbital structure
of the Kuiper belt, including several categories of objects inside and outside of resonances with Neptune, 
emerged as a result of Neptune's migration into an outer planetesimal disk. An outstanding problem with
the existing migration models is that they invariably predict excessively large resonant populations, 
while observations show that the non-resonant orbits are in fact common (e.g., the main belt population 
is $\simeq$2-4 times larger than Plutinos in the 3:2 resonance). Here we show that this problem can resolved 
if it is assumed that Neptune's migration was {\it grainy}, as expected from scattering encounters of 
Neptune with massive planetesimals. The grainy migration acts to destabilize resonant bodies with large 
libration amplitudes, a fraction of which ends up on stable non-resonant orbits. Thus, the non-resonant--to--resonant 
ratio obtained with the grainy migration is higher, up to $\sim$10 times higher for the range of parameters investigated here, 
than in a model with smooth migration. In addition, the grainy migration leads to a narrower distribution of the 
libration amplitudes in the 3:2 resonance. The best fit to observations is obtained when it is assumed that the 
outer planetesimal disk below 30 AU contained 1000-4000 Plutos, or $\sim$1000 bodies twice as massive as Pluto.
We find that the probability for an outer disk object to end up on a stable orbit in the Kuiper belt is 
$\sim$$10^{-3}$. Together, these results are consistent with having only two (known) Pluto-mass bodies in 
the Kuiper belt today (Pluto and Eris). We estimate that the combined mass of Pluto-class objects in the original disk 
was $\sim$2-8 Earth masses ($M_{\rm Earth}$), which represents 10-40\% of the estimated disk mass ($M_{\rm disk}\simeq20$ 
$M_{\rm Earth}$). This constraint can be used to better understand the accretion processes in the outer Solar 
System.
\end{abstract}
\section{Introduction}
Studies of Kuiper belt dynamics first considered the effects of outward migration of Neptune (Fern\'andez \& Ip 1984) that can explain 
the prominent populations of Kuiper Belt Objects (KBOs) in major resonances (Malhotra 1993, 1995; Hahn \& Malhotra 1999, 2005; Chiang \& 
Jordan 2002; Chiang et al. 2003; Levison \& Morbidelli 2003; Gomes 2003; Murray-Clay \& Chiang 2005, 2006). Adding to that, 
Petit et al. (1999) invoked the dynamical effect of large planetesimals scattered from the Neptune region and showed that it 
can explain the general depletion and excitation of the belt. With the advent 
of the notion that the early Solar System may have suffered a dynamical instability (Thommes et al. 1999, Tsiganis et al. 2005, 
Morbidelli et al. 2007), the focus broadened, with the more recent theories invoking a transient phase with an eccentric 
orbit of Neptune (Levison et al. 2008, Morbidelli et al. 2008, Batygin et al. 2011, Wolff~et~al.~2012, Dawson \& Murray-Clay 2012). 
The consensus emerging from these studies is that the Hot Classical (hereafter HC), resonant, scattered and detached populations 
(see Gladman et al. 2008 for the definition of these groups), formed in a massive planetesimal disk at $\lesssim$30 AU, and 
were dynamically scattered onto their current orbits by migrating (and possibly eccentric) Neptune (Levison et al. 2008, 
Morbidelli et al. 2008, Dawson \& Murray-Clay 2012). The Cold Classicals (CCs), on the other hand, probably formed at $>$40~AU 
and survived Neptune's early `wild days' relatively unharmed (Kavelaars et al. 2009, Batygin et al. 2011, Wolff et al. 2012). 

\begin{figure}
\epsscale{0.6}
\plotone{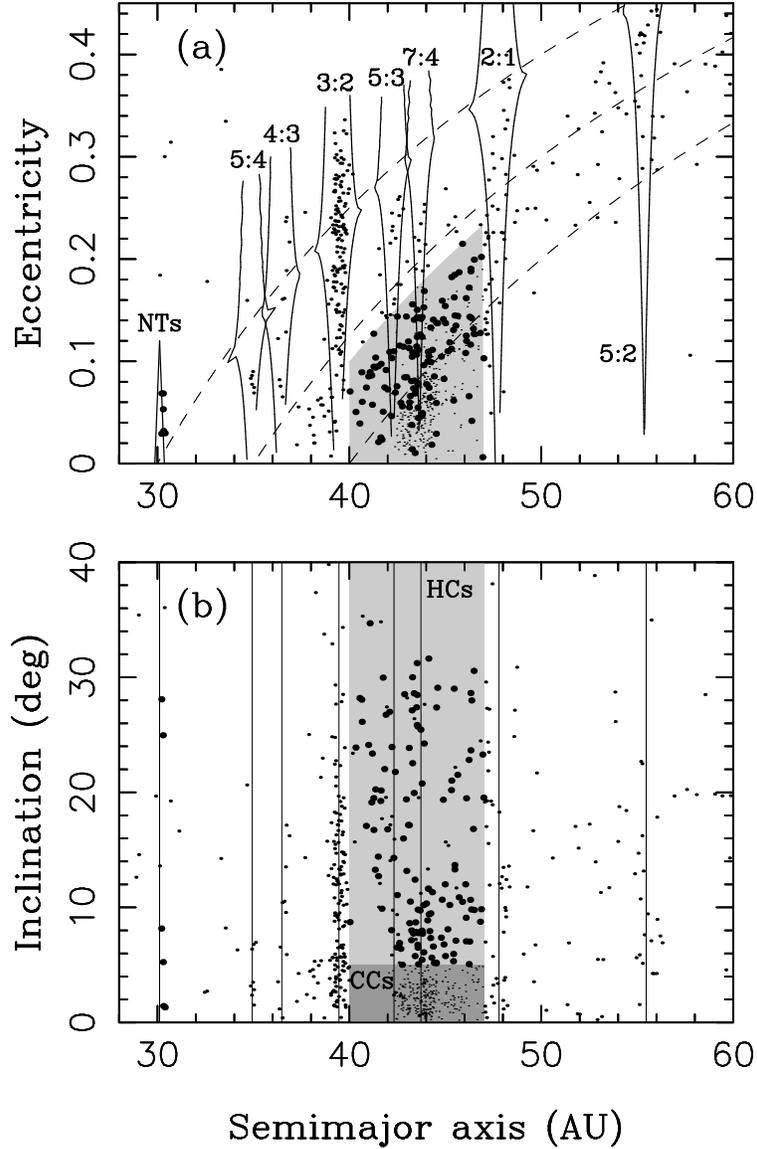}
\caption{The orbital elements of KBOs observed in three or more oppositions. Various dynamical classes 
are highlighted. The HCs with $i>5^\circ$ and Neptune Trojans are denoted by larger dots, and the CCs are denoted 
by smaller dots. Note the wide inclination 
distribution of the HCs in panel (b) with inclinations reaching above $\simeq$$30^\circ$. The solid lines 
in panel (a) follow the borders of important mean motion resonances. For Neptune Trojans, we show an approximate 
location of stable librations. The low-inclination orbits
with $40<a<42$ AU are unstable due to an overlap of the secular resonances $\nu_7$ and $\nu_{8}$ (Kn\v{e}\v{z}evi\'c et al. 1991, 
Duncan et al. 1995).}
\label{fig1}
\end{figure}

In our previous work, we considered two unexplained features of the Kuiper belt. First, we examined the wide inclination distribution 
of the HCs and resonant populations (Fig.~\ref{fig1}; Nesvorn\'y 2015a). We found that this is key to understanding the emergence 
of the Kuiper belt. Specifically, the inclination distribution implies that Neptune's migration must have been long range (Neptune 
starting below $\simeq$25 AU), and slow (exponential e-folding timescale $\tau \gtrsim 10$ Myr). This is because Neptune needs 
to be given sufficient time to raise the orbital inclinations by close encounters with the disk objects. Second, we showed that 
the concentration of CCs near 44 AU, known as the Kuiper belt kernel (Petit et al. 2011), can be explained if Neptune's otherwise 
smooth migration was interrupted by a discontinuous change in Neptune's orbit when Neptune reached $\simeq$28 AU (Nesvorn\'y 2015b). 
The kernel forms in this model as bodies previously collected in Neptune's 2:1 resonance are released at $\simeq$44 AU when Neptune 
jumps. Taken together, these results provide support for the planetary migration/instability model developed in Nesvorn\'y \& Morbidelli 
(2012), where Neptune slowly migrates from $\lesssim$25 AU to $\simeq$28 AU, jumps by $\simeq$0.5 AU by being scattered off of 
another planetary body during the instability, and then continues migrating to the original edge of the massive planetesimal 
disk at 30 AU (see also Gladman et al. 2012; discussion in Section 13).

\begin{figure}
\epsscale{0.6}
\plotone{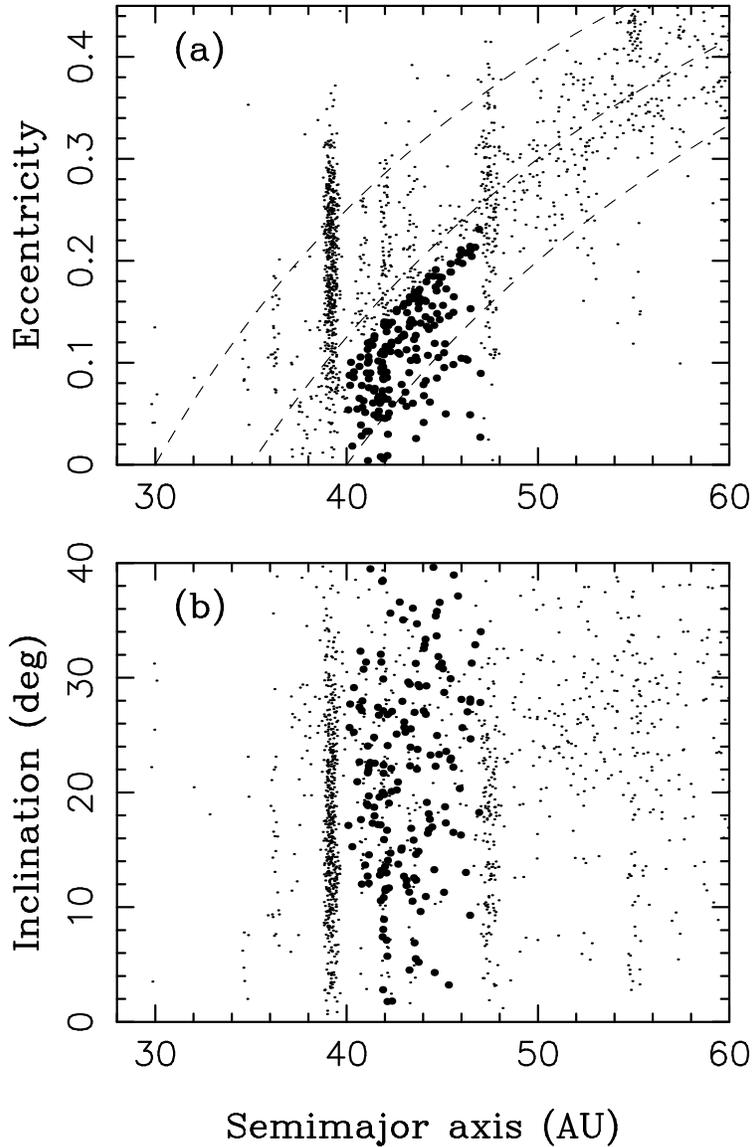}
\caption{The orbital elements of bodies captured in the Kuiper belt in a model with smooth migration, 
$a_{\rm N,0}=24$~AU and $\tau=30$~Myr (from Nesvorn\'y 2015a). The bodies captured on orbits in the main belt 
region are denoted by larger symbols. Note the very large population of Plutinos ($a\simeq39.5$ AU)
obtained in this model. There are nearly three times as many Plutinos as the HCs in the plot, while observations 
indicate that in reality there should be $\simeq$2-4 times {\it fewer} Plutinos than the HCs. This clearly 
illustrates the resonance overpopulation problem.}
\label{smooth}
\end{figure}

\begin{figure}
\epsscale{0.6}
\plotone{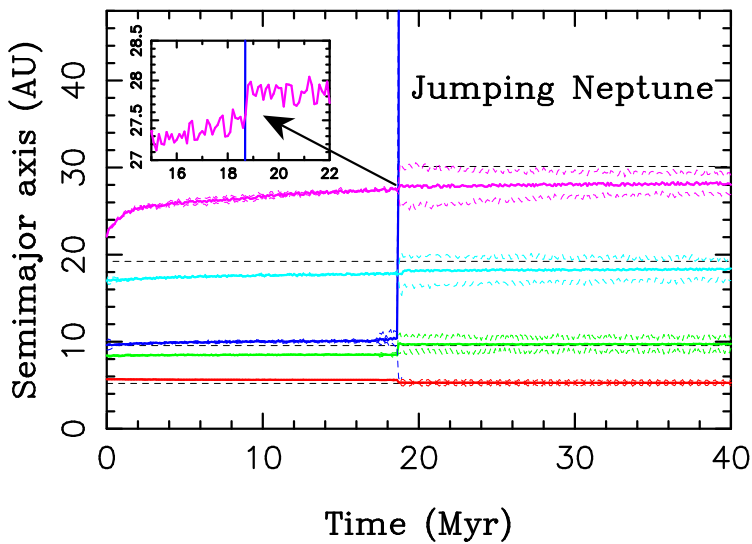}
\caption{The orbit histories of the giant planets in an instability simulation from NM12. In this example, the fifth giant 
planet was initially placed on an orbit between Saturn and Uranus and was given a mass equal to the Neptune mass. Ten 
thousand particles, representing the outer planetesimal disk, were distributed with the semimajor axis $23.5<a<29$ AU, 
surface density $\Sigma=1/a$, and low eccentricity and low inclination. With the total disk mass $M_{\rm disk}=15$ $M_{\rm Earth}$, 
each disk particle has $\simeq$0.75 Pluto mass. The plot shows the semimajor axes (solid lines), and perihelion and aphelion 
distances (thin dashed lines) of each planet's orbit in a time frame $\pm20$ Myr around the instability. Neptune migrates into 
the outer disk during the first stage of the simulation. It reaches $\simeq$27.5 AU when the instability happens ($t\simeq18.3$ Myr). 
During the instability, Neptune has a close encounter with the fifth planet and its semimajor axis jumps by 
$\simeq$0.4 AU outward (see the inset). The fifth planet is subsequently ejected from the solar system by Jupiter. Neptune's 
migration after the instability can be approximated with the e-folding timescale $\tau_2=50$ Myr. The effective $\tau_2$ becomes 
longer ($\tau_2 \gtrsim 100$ Myr) at later times. The final orbits of the four remaining planets are a good match to those 
in the present Solar System (thin dashed lines).}
\label{case9}
\end{figure}

Nesvorn\'y (2015a) pointed out an outstanding problem with previous simulations of the Kuiper belt formation (e.g. Hahn \& Malhotra 
2005, Levison et al. 2008), including their own model. They called it the {\it resonance overpopulation problem}. This problem 
arises when the number of resonant objects in the 3:2 resonance, $N_{\rm 3:2}$, 
is compared to the number of HCs ($N_{\rm HC}$). According to observations, Plutinos in the 
3:2 resonance are $\simeq$2-4 times less numerous than the HCs (Gladman et al. 2012, Adams et al. 2014). Thus, $N_{\rm HC}/N_{\rm 3:2} \simeq 2$-4. 
The populations in the 2:1 and 5:2 resonances are probably somewhat smaller than $N_{\rm 3:2}$. In contrast, the simulations of Nesvorn\'y 
(2015a), where Neptune was slowly and {\it smoothly} migrated from $a_{\rm N,0}<25$ AU to 30 AU, give $N_{\rm HC}/N_{\rm 3:2}\sim 0.2$-0.5 
(Fig. \ref{smooth}).\footnote{Nesvorn\'y (2015a) reported a few special cases with a smooth migration  
where $N_{\rm HC}/N_{\rm 3:2} \gtrsim 1$, possibly in better agreement with observations. These cases correspond to very long migration timescales 
($\tau \geq 100$~Myr). While this could help to resolve the resonance overpopulation problem, these very long migration timescales lead 
to several problems elsewhere. For example, the inclination distribution of HCs and Plutinos in the 3:2 resonance obtained with $\tau \geq 100$ Myr 
is wider than indicated by observations, and the total implantation efficiency into the Kuiper belt is $\simeq5\times10^{-3}$, which is 
probably excessive. We therefore believe that the very long migration timescales do not provide a viable solution of the resonance 
overpopulation problem.} A possible solution to this problem, suggested in Nesvorn\'y (2015a), is the {\it jumping Neptune model}, in which 
Neptune radially jumps by being scattered off of another planet (Fig. \ref{case9}). While the jumping Neptune model was primarily motivated by
the formation of the Kuiper belt kernel (Petit et al. 2011, Nesvorn\'y 2015b), it can also help to reduce the resonant populations, because 
bodies captured in resonances before Neptune's jump are released when Neptune jumps, and thus do not contribute to the final statistics. 

Here we conduct numerical simulations of the jumping Neptune model and find that Neptune's jump helps, but is not sufficient 
to reconcile the model with observations. We therefore investigate other solutions to the resonance overpopulation problem. We find that 
the problem can be resolved if Neptune's migration was {\it grainy} due to a presence of Pluto-class objects in the planetesimal disk 
that was driving the planetary migration. The principal difference between the smooth and grainy migration modes is that in the latter case 
Neptune's resonances exhibit a random walk in the semimajor axis (in addition to the smooth radial drift). This acts to reduce the resonant
populations, because resonant orbits with large libration amplitudes can become unstable. At the same time, it helps to increase the HC 
population, because orbits evolve from Neptune's resonances onto stable non-resonant orbits more easily than in the smooth case. Specifically, we find 
that $N_{\rm HC}/N_{\rm 3:2} \sim 2$-4 is obtained in the grainy migration model if the planetesimal disk is assumed to have contained 
$\sim$1000-4000 Plutos, or $\sim$1000 bodies twice as massive as Pluto. Sections 2 and 3 describe our method and results, respectively. 
Broader implications of this work are discussed in Section 4. 
\section{The Integration Method}
The integration method with smooth migration of Neptune is explained in Section 2.1. The migration parameters were chosen
to match the orbital evolution of planets obtained in the self-consistent simulations of the planetary instability/migration
in Nesvorn\'y \& Morbidelli (2012; hereafter NM12). The initial distribution of disk particles is defined in Section~2.2.
Then, in Section 2.3, we introduce massive objects in the outer disk and let Neptune react to individual scattering events. 
Section 2.4 explains how we used the Canada-France Ecliptic Plane Survey (CFEPS) detection simulator to compare our modeling
results with observations.  
\subsection{Smooth Migration}
Our numerical integrations consist of tracking the orbits of four giant planets (Jupiter to Neptune) and a large number 
of test particles representing the outer planetesimal disk. To set up an integration, Jupiter and Saturn were placed on their 
current orbits. Uranus and Neptune were placed on inside of their current orbits and were migrated outward. The initial 
semimajor axis $a_{\rm N,0}$, eccentricity $e_{\rm N,0}$, and inclination $i_{\rm N,0}$ define Neptune's orbit before the main 
stage of migration/instability. In most of our simulations we used $a_{\rm N,0}=24$~AU, because the wide inclination 
distribution of HCs and resonant populations requires that Neptune's migration was long range ($a_{\rm N,0}\lesssim25$ 
AU; Nesvorn\'y 2015a). We also set $e_{\rm N,0}=0$ and $i_{\rm N,0}=0$. All inclination values reported in this article are
referred to the invariant plane of the Solar System. 

The {\tt swift\_rmvs4} code (Levison \& Duncan 1994) was used to follow the evolution of planets (and massless disk 
particles; see Sect 2.2). 
The code was modified to include fictitious forces that mimic the radial migration and damping of planetary orbits. These forces were 
parametrized by the exponential e-folding timescales, $\tau_a$, $\tau_e$ and $\tau_i$, where $\tau_a$ controls the radial migration 
rate, and $\tau_e$ and $\tau_i$ control the damping rate of $e$ and $i$. Specifically, the semimajor axis of Neptune
changes from $a_{\rm N,0}$ to its current average of $a_{\rm N,c}=30.11$ AU as
\begin{equation}
a_{\rm N}(t)=a_{\rm N,c}+(a_{\rm N,0}-a_{\rm N,c}) \exp(-t/\tau_a)\ ,
\end{equation}
and the eccentricity and inclination of Neptune evolve according to
\begin{equation}
e_{\rm N}(t)=e_{\rm N,0} \exp(-t/\tau_e)\ {\rm and}\ i_{\rm N}(t)=i_{\rm N,0} \exp(-t/\tau_i)\ .
\end{equation}
The expressions for $e_{\rm N}(t)$ and $i_{\rm N}(t)$ differ from those used in Morbidelli et al. (2014), where
the damping rate $({\rm d}e_{\rm N}/{\rm d}t)/e_{\rm N}$
was chosen to be proportional to $\exp(-t/\tau_i)$. Here we set $\tau_a \sim \tau_e \sim \tau_i$ $(=\tau_1)$, 
because such roughly comparable timescales were suggested by previous work.

The numerical integrations of the first migration stage were stopped when Neptune reached $a_{\rm N,1}\simeq28$ AU. Then,
to approximate the effect of planetary encounters during the instability (NM12, Nesvorn\'y 2015b), we applied a discontinuous 
change of Neptune's semimajor axis and eccentricity, $\Delta a_{\rm N}$ and $\Delta e_{\rm N}$. Motivated  by the NM12 results
(see Fig.~\ref{case9}), we set $\Delta a_{\rm N}=0$ or 0.5 AU, and $\Delta e_{\rm N}=0$, 0.05 or 0.1. The purpose of $\Delta 
a_{\rm N}=\Delta e_{\rm N}=0$ is to have a reference case for comparison purposes. We use $\Delta a_{\rm N}=0.5$ AU, because
Nesvorn\'y (2015b) showed that this jump size would be needed to explain the Kuiper belt kernel. Note that the resonant
objects are released from resonances with $\Delta a_{\rm N}=0.5$ AU, because the typical resonance width is just smaller
than the jump size. No change was applied to the orbital inclination of Neptune, because a small inclination change 
should not critically affect the processes studied here. 

The second migration stage starts with Neptune having the semimajor axis $a_{\rm N,2}=a_{\rm N,1}+\Delta a_{\rm N}$. 
We apply the {\tt swift\_rmvs4} code, and migrate the semimajor axis (and damp the eccentricity) on an e-folding 
timescale $\tau_2$. The migration amplitude was adjusted such that the final semimajor axis of Neptune ended to be within 
0.05~AU of its current mean $a_{\rm N,c}=30.11$~AU, and the orbital period ratio, $P_{\rm N}/P_{\rm U}$, where $P_{\rm N}$ 
and $P_{\rm U}$ are the orbital periods of Neptune and Uranus, ended within 0.5\% of its current value 
($P_{\rm N}/P_{\rm U}=1.96$). A strict control over the final orbits of planets is important, because it guarantees that 
the mean motion and secular resonances reach their present positions. 

As for the specific values of $\tau_1$ and $\tau_2$ used in our model, we found from NM12 that the orbital behavior 
of Neptune can be approximated by $\tau_1\simeq10$ Myr and $\tau_2\simeq30$~Myr for a disk mass $M_{\rm disk}=20$ $M_{\rm Earth}$, 
and $\tau_1\simeq20$ Myr and $\tau_2\simeq50$ Myr for $M_{\rm disk}=15$~$M_{\rm Earth}$. Slower migration rates 
are possible for lower disk masses. Moreover, we found from NM12 that the real migration is not precisely exponential with the effective 
$\tau_2$ being longer than the values quoted above during the very late migration stages ($\tau_2 \gtrsim 100$ Myr). This is 
consistent with constraints from Saturn's obliquity, which was presumably exited by late capture in a spin-orbit 
resonance (Vokrouhlick\'y \& Nesvorn\'y 2015; see Ward \& Hamilton 2004 and Hamilton \& Ward 2004 for the original work 
that proposed the spin-orbit resonance as the means of exciting Saturn's obliquity). 
Much shorter migration timescales than those quoted above do not apply, 
because they would violate constraints from the wide inclination distribution of HCs and resonant populations (Nesvorn\'y 
2015a). Here we therefore used $\tau_1=10$~Myr or 30 Myr, and $\tau_2=30$ or 100 Myr. These cases should bracket the range
of possible migration timescales. 

\subsection{Planetesimal Disk Properties} 

The planetesimal disk was divided into two parts. The inner part of the disk, from just outside Neptune's initial orbit to 
$r_{\rm edge}$, was assumed to be massive. We used $r_{\rm edge}=28$~AU or 30 AU, because our previous simulations in NM12 showed 
that the massive disk's edge must be at 28-30 AU for Neptune to stop at $\simeq$30 AU. If the edge of the massive disk were
at $>$30~AU, Neptune would continue migrating past 30 AU (Gomes et al. 2004). The solar nebula could have become truncated, for 
example, by photoevaporation from the UV and FUV irradiation by background stars in a cluster (e.g., Adams 2010; see also 
discussion in Petit et al. 2011). In fact,
a recent study of the dynamics of planetesimals embedded in a gas disk suggested that the solar nebula was truncated
(or else it would act to produce very high orbital inclinations, $i>40^\circ$, in the Kuiper belt; Kretke et al. 2012).
The estimated mass of the planetesimal disk below 30 AU is $M_{\rm disk}\simeq20$~$M_{\rm Earth}$ (NM12). As shown in Levison et al. (2008), 
the massive disk is the main source of HCs, Plutinos and other resonant populations. It therefore has a crucial 
importance for the resonance overpopulation problem considered here.

The planetesimal disk probably had a low mass extension reaching from 30 AU to at least $\simeq$45 AU. 
The low mass extension of the disk beyond 30 AU is presumably the source of CCs (Batygin et al. 2011, Wolff et al. 2012, Nesvorn\'y 2015b). 
It is needed to explain why the CCs have several unique physical and orbital properties (see Section 3.4). The disk extension  
should not substantially contribute to the present populations of the hot and resonant KBOs (Nesvorn\'y 2015b)\footnote{With 
a possible exception of the 2:1 resonance, which sweeps through the low mass extension of the disk during Neptune's migration, 
and can capture and retain an important population of low-inclination orbits. The orbital inclinations of known KBOs in the 2:1 
resonance may hint on this, but better statistics will be  needed to establish things more firmly.}, because 
the orbital inclinations of bodies native to $a>40$ AU remain small during Neptune's migration. Here we therefore do not initially
consider the disk extension, and return to it only in Section 3.4, where we test whether a grainy migration is consistent with the 
formation of the Kuiper belt kernel.

Each of our simulations included one million disk particles distributed from outside Neptune's initial orbit to 
$r_{\rm edge}$. The radial profile was set such that the disk surface density $\Sigma \propto 1/r$, where $r$ is the heliocentric 
distance. A large number of disk particles was needed because the implantation probability in different parts of the 
Kuiper belt is expected to be $\sim$$10^{-3}$-$10^{-4}$. With $10^6$ disk particles initially, this yields $\sim$100-1000 
implanted particles, and allows us to perform a detailed comparison of the model results with observations.
The disk particles were assumed to be massless such that their gravity does not interfere with the migration/damping routines. 
This means that the precession frequencies of planets are not affected by the disk in our simulations, while in reality 
they were. This is an important approximation (Batygin et al. 2011). The direct gravitational effects of the fifth planet 
on the disk planetesimals were ignored (see discussion at the end of Section 4). 
These effects could be especially important for the CCs (Batygin et al. 2012).

An additional uncertain parameter concerns the dynamical structure of the planetesimal disk. It is typically 
assumed that the disk was dynamically cold with orbital eccentricities $e\lesssim0.1$ and orbital inclinations $i\lesssim10^\circ$. 
Some dynamical excitation could have been supplied by scattering off of Pluto-sized and larger objects that presumably 
formed in the disk (Stern \& Colwell 1997, Kenyon et al. 2008). The magnitude of the initial excitation is uncertain, 
because it depends on several unknown parameters (e.g., the number of large objects in the disk). The initial eccentricities 
and initial inclinations of disk particles in our simulations were distributed according to the Rayleigh distribution 
with $\sigma_e=0.1$ and $\sigma_i=0.05$, where $\sigma$ is the usual scale parameter of the Rayleigh distribution
(the mean of the Rayleigh distribution is equal to $\sqrt{\pi/2}\sigma$).

\subsection{Grainy Migration}

We developed an analytic method to represent the jitter that Neptune's semimajor axis experiences due to close
encounters with massive planetesimals. The method has the flexibility to use any smooth migration history of Neptune 
on the input, include a certain number of the massive planetesimals in the original disk, and generate a new migration history 
where the random element due to massive planetesimal encounters is included. This approach is useful, because we can
easily control how grainy the migration is, while preserving the global orbital evolution of planets from the smooth
simulations. We then proceed to test how the simulation results depend on various parameters, such as the number and 
mass of the massive planetesimals in the original disk. 

We start with a specific migration run in which Neptune's semimajor axis evolves smoothly, except
for a possible jump by $\Delta a_{\rm N}$ due to an encounter with another planet. The migration 
parameters,  namely $a_{\rm N,0}$, $a_{\rm N,1}$, $\Delta a_{\rm N}$, $\tau_1$ and $\tau_2$ (Section 2.1), 
are specified at this point. As mentioned above, each run also includes $10^6$ test particles that represent 
the disk planetesimals. We first scan through the simulation output in small steps $\Delta t$, and extract 
the orbit of Neptune and the orbital distribution of disk planetesimals at each step. We then apply
the \"Opik-type collision probability code (Bottke et~al. 1994; see also Greenberg 1982) to calculate how 
many encounters between Neptune and planetesimals happen for encounter distances $r<R$, where $R$ is 
some threshold. The gravitational focusing by Neptune is taken into account in this calculation.

The goal is to find how the number of encounters with $r<R$ depends on $R$. We find that for small values of $R$ 
this dependence is linear (while a quadratic dependence would be expected without gravitational focusing). 
This can be understood from the following expression for the impact parameter: 
\begin{equation}
 b^2_{\rm max} = R^2 \left[1 + \frac{R_{\rm N}}{R}\left(\frac{v_{\rm esc}}{v_\infty}\right)^2
  \right] \label{eq:bq}
\end{equation}
(e.g., Bertotti et~al. 2003), where $R_{\rm N}=24,622$ km is Neptune's mean radius, $v_{\rm esc}\simeq 23.5$~km~s$^{-1}$ 
is the escape velocity from Neptune's cloudtops, and $v_\infty$ is the encounter speed ``at infinity''. 
Parameter $b_{\rm max}(R)$ is the maximal impact parameter value for which the minimal encounter distance is 
lower than specified $R$, when $v_\infty$ is fixed. 
Since $v_\infty \simeq 1$-2~km s$^{-1}$, and thus $v_\infty \ll v_{\rm esc}$, the second term in Eq. (\ref{eq:bq}) 
is greater than 1 for all encounters with $R<R^*$, where $R^*=R_{\rm N}(v_{\rm esc}/v_\infty)^2$. The number of encounters 
with $r<R$ in this regime is therefore proportional 
to $R$. For $R>R^*$, on the other hand, the first term in Eq. (\ref{eq:bq}) prevails, and the number of 
encounters with $r<R$ becomes proportional to $R^2$. In practice, we find it satisfactory to neglect the effect
of distant encounters, because the distant encounters do not (individually) induce any significant changes of 
Neptune's semimajor axis. We therefore only consider encounters with $r<R^*$, where the scaling is linear.    
Note that $R^* > 140\ R_{\rm N}$ for $v_\infty < 2$~km s$^{-1}$. 

The \"Opik code gives us the number of planetesimals having encounters with Neptune, $n(R,t; \Delta t)$,  
as a function of the distance $R$, time $t$, and time interval $\Delta t$. Obviously, $n(R,t; \Delta t) \propto 
\Delta t$ for the small intervals used here ($\Delta t=10^3$ yr). The time profile of the number of 
encounters depends on the timescale of the planetesimal disk dispersal, which in turn is related to Neptune's migration 
speed. We find from our simulations that $n(R,t; \Delta t)\propto R\,t^{-\alpha}$ with some exponent $\alpha$. For example, in the simulation with 
$\tau_1=30$~My and $\tau_2=100$~My, we have $\alpha \simeq 1.1$-1.2. Then, if we approximate $n(R,t; \Delta t)= 
C\,R\,t^{-\alpha}$, the constant $C$ will depend on the initial number of massive planetesimals in the disk
($N_{\rm mp}$). Here we consider $N_{\rm mp}=1000$, 2000 and 4000, and rescale the original $n(R,t; \Delta t)$ 
obtained with $10^6$ disk particles to $N_{\rm mp}$. This gives us an approximate record of the history 
of close encounters between the massive planetesimals and Neptune. 

Equipped with the calibrated $n(R,t; \Delta t)$ function, we sequentially consider each $\Delta t$ interval. 
With $\Delta t=10^3$ yr, $n(R,t; \Delta t)$ is always much smaller than unity. In each step, we generate 
a random variate $X$ with a uniform distribution between $0$ and $1$. We then set $n(R,t; \Delta t)=X$
and solve for $R=R(t,X; \Delta t)$. If $R_{\rm N} < R < R_{\rm max}$, where $R_{\rm max}<R^*$ is the maximum 
encounter distance considered here (in practice we set $R_{\rm max}=R_{\rm Hill}/10$, where $R_{\rm Hill}=$ is
the Hill radius), the encounter is recorded, and we proceed with the next timestep $\Delta t$. 
In the end, the method allows us to generate an encounter sequence that closely approximates the reality. 
Note that for the typical encounter speeds considered here it only takes up to several months to cross $R_{\rm max}$. 

We proceed by computing the effect of individual encounters on the semimajor axis of Neptune. This is done using
a hyperbolic approximation of the planetesimal's trajectory relative to Neptune. The hyperbolic approximation 
is adequate for deep encounters considered here, and in a regime when the encounter duration is much 
shorter than the orbital period of Neptune. 
The deflection angle $\theta$ of the asymptotes of the hyperbola describing the planetesimal encounter trajectory 
can be obtained from $R$ and $v_\infty$. Specifically, introducing the half-angle $\theta_{1/2}=\case{1}{2}\,\theta$, 
we find (e.g., Bertotti et~al., 2003)
\begin{equation}
 \sin\theta_{1/2} = \left[1 + 2\, \frac{R}{R_{\rm N}}\left(\frac{v_\infty}{v_{\rm esc}}
  \right)^2\right]^{-1}\; . \label{eq:th}
\end{equation}
Expressed in the inertial frame, the change of Neptune's velocity vector is
\begin{equation}
 \delta {\bf V} \simeq 2\, \frac{m}{M_{\rm N}}\,v_\infty \sin\theta_{1/2} \left[
  \left({\bf e}_1\times{\bf e}_2\right)\cos\theta_{1/2} -
  {\bf e}_2 \,\sin\theta_{1/2}\right]\; , \label{eq:dvpl}
\end{equation}
where $M_{\rm N}$ and $m$ are Neptune's and planetesimal masses, ${\bf e}_1$ is a unit vector along the angular 
momentum vector of planetesimal's planetocentric trajectory, and ${\bf e}_2$ is a unit vector along the incoming 
trajectory of the planetesimal (relative to the planet). Since we do not propagate the information about 
${\bf e}_1$ and ${\bf e}_2$ from the original simulation, we assume that ${\bf e}_1$ and ${\bf e}_2$ are randomly 
(isotropically) oriented in space (obviously, these vectors are perpendicular to each other). This should be 
a reasonable approximation in a situation when the disk of planetesimals is dispersed by Neptune.  
 
Finally, we use the Gauss equations to compute the orbital effect of $\delta {\bf V}$ on Neptune's orbit. The 
eccentricity and inclination changes are neglected. In the approximation of a near-circular orbit of Neptune, we obtain
\begin{equation}
 \frac{\delta a_{\rm N}}{a_{\rm N}}\simeq 2\,\frac{\delta V_\parallel}{V_{\rm N}}\; ,
  \label{eq:da}
\end{equation}
where $\delta a_{\rm N}$ is the change of Neptune's semimajor axis, $\delta V_\parallel$ is a projection of $\delta {\bf V}$ 
onto the transverse direction of Neptune's orbital motion, and $V_{\rm N}$ is Neptune's orbital speed. 
To compute $\delta V_\parallel$ from $\delta {\bf V}$, we assume that $\delta {\bf V}$ has a random orientation.
The change $\delta a_{\rm N}$ is computed for all encounters with massive planetesimals and recorded in a file. In 
a simulation, we then use the modified {\tt swift\_rmvs4} code with smooth migration (Section 2.1), 
and apply $\delta a_{\rm N}$ every time that is recorded in the encounter file. 

\begin{figure}
\epsscale{0.7}
\plotone{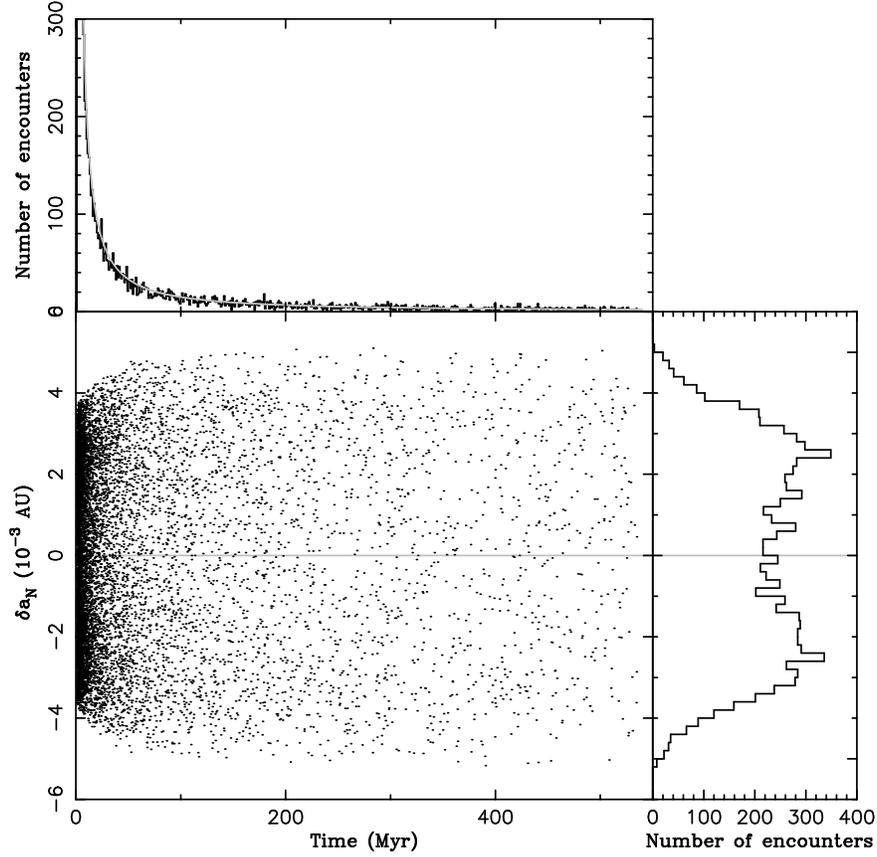}
\caption{The sequence of Neptune's semimajor axis changes, $\delta a_{\rm N}$, due to massive planetesimal encounters.
This sequence was generated for a case with $\tau_1=30$~Myr, $\tau_2=100$~Myr, and $N_{\rm mp}=1000$. Each massive planetesimal 
was assumed to have one Pluto mass. The top panel shows the number of encounters per $1$~Myr as a function of time.
The gray line in the top panel is a power law function, $t^{\alpha}$ with $\alpha=-1.15$, that provides an excellent match to 
the decreasing profile of the number of encounters with time. The central panel shows the $\delta a_{\rm N}$ values produced 
by individual encounters. The histogram on the right is the distribution of $\delta a_{\rm N}$.
\label{fig_enc}}
\end{figure}

Figure~\ref{fig_enc} shows an example of the sequence of $\delta a_{\rm N}$ in the case with $\tau_1=30$~Myr, 
$\tau_2=100$~Myr, and $N_{\rm mp}=1000$, where each massive planetesimal was assumed to have Pluto's mass.
The total number of encounters recorded in this case is $\sim$$10^4$, implying that a massive planetesimal had
on average $\sim$$10$ encounters with $r < R_{\rm max}=R_{\rm Hill}/10$. As expected, most encounters happen during 
the initial migration stages when the planetesimal disk is still massive. The slight preference for somewhat larger 
range of the $\delta a_{\rm N}$ values at late epochs reflects the outward migration of Neptune's orbit. The root mean square 
of $\delta a_{\rm N}/a_{\rm N}$ is of the order of $\sim (m/M)\,(v_\infty/V_{\rm N})\sim 10^{-4}$, as expected from Eqs.~(\ref{eq:dvpl}) 
and (\ref{eq:da}). By design, the $\delta a_{\rm N}$ distribution has a zero mean, while in reality the distribution
should be skewed toward positive values, because the massive planetesimals contribute to Neptune's outward migration.

\begin{figure}
\epsscale{0.8}
\plotone{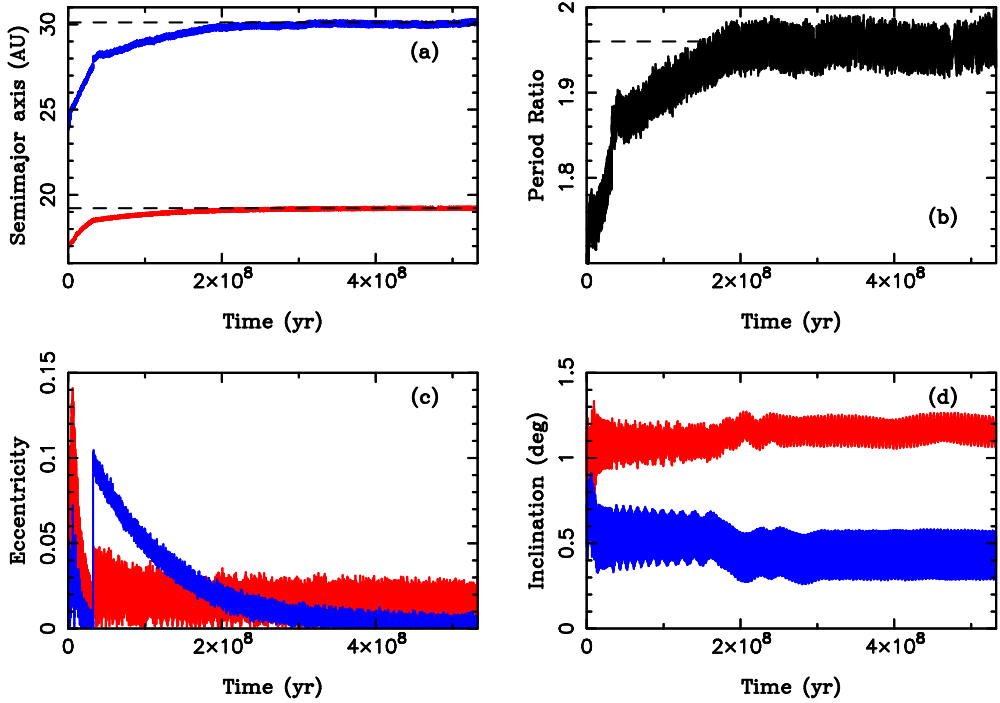}
\caption{The orbital histories of the outer planets in a simulation with $\tau_1=30$ Myr, $\tau_2=100$~Myr,
$a_{\rm N,1}=27.8$ AU, $\Delta a_{\rm N} =0.5$ AU, $\Delta e_{\rm N} = 0.1$. Here we assumed that the outer disk contained
1000 massive planetesimals each with mass $M_{\rm mp}=2\ M_{\rm Pluto}$, and applied the method described in Section 2.3 to mimic a
grainy migration that would result from the interaction of Neptune with these massive objects. Neptune's jump happens at 
$t=32.5$ Myr in this simulation. Panel (b) shows the orbital period ratio $P_{\rm N}/P_{\rm U}$. The horizontal dashed lines in 
panels (a) and (b) correspond to the present values of the semimajor axes of Uranus and Neptune.}
\label{mig}
\end{figure}

Figure \ref{mig} shows an example of grainy migration of Neptune produced by the method described in this section. 
The effects of encounters are clearly visible in panel (b), where the ratio of orbital periods, $P_{\rm N}/P_{\rm U}$,
shows an irregular pattern. During the late stages, when the smooth migration nearly stalls, the random effect 
of encounters with massive planetesimals can cause Neptune to move slightly inward during some time intervals. 
This may be important for Neptune Trojans, because their stability sensitively depends on the maximal $P_{\rm N}/P_{\rm U}$
reached during the planetary migration (Gomes \& Nesvorn\'y 2015).
\subsection{The CFEPS Detection Simulator} 
We used the CFEPS detection simulator (Kavelaars et al. 2009) to compare the orbital distributions obtained 
in our simulations with observations. CFEPS is one of the largest Kuiper belt surveys with published characterization
(currently 169 objects; Petit et al. 2011). The simulator was developed by the CFEPS team to aid the interpretation of their 
observations. Given intrinsic orbital and magnitude distributions, the CFEPS simulator returns a sample of objects that would 
have been detected by the survey, accounting for flux biases, pointing history, rate cuts and object leakage (Kavelaars et al. 2009).   
In the present work, we input our model populations in the simulator to compute the detection statistics. 
We then compare the orbital distribution of the detected objects with the actual CFEPS detections using the 
Kolmogorov-Smirnov (K-S) test (Press et al. 1992). 

This is done as follows. The CFEPS simulator takes as an input: (1) the orbital element distribution from 
our numerical model, and (2) an assumed absolute magnitude ($H$) distribution. As for (1), the input orbital 
distribution was produced by a short integration starting from the final model state of the Kuiper belt. 
The orbital elements of each object were recorded at 100 yr intervals during this integration until the 
total number of the recorded data points reached $\simeq$$10^5$. Each data point was then treated as an independent 
observational target. We rotated the reference system such that the orbital phase of Neptune in each time 
frame corresponded to its ecliptic coordinates at the epoch of CFEPS observations. This procedure guaranteed 
that the sky positions of bodies in Neptune's resonances were correctly distributed relative to the 
pointing direction of the CFEPS frames.

The magnitude distribution was taken from Fraser et al. (2014). It was assumed to be described by a broken 
power law with 
$N(H)\,{\rm d}H=10^{\alpha_1(H-H_0)}\,{\rm d}H$ for $H<H_{\rm B}$ 
and  
$N(H)\,{\rm d}H=10^{\alpha_2(H-H_0)+(\alpha_1-\alpha_2)(H_{\rm B}-H_0)}\,{\rm d}H$ for $H>H_{\rm B}$,
where $\alpha_1$ and $\alpha_2$ are the power-law slopes for objects brighter and fainter than 
the transition, or break magnitude $H_{\rm B}$, and $H_0$ is a normalization constant.  Fraser et al. (2014) 
found that $\alpha_1\simeq0.9$, $\alpha_2\simeq0.2$ and $H_{\rm B}\simeq8$ for the HCs. In the context of a model where the 
HCs formed at $<$30 AU, and were implanted into the Kuiper belt by a size-independent process (our 
integrations do not have any size-dependent component), the HC magnitude distribution should be shared 
by all populations that originated from $<$30 AU (Morbidelli et al. 2009 , Fraser et al. 2014). We varied 
the parameters of the input magnitude distribution to understand the sensitivity of the results to
various assumptions. We found that small variations of $\alpha_1$, $\alpha_2$ and $H_{\rm B}$ within
the uncertainties given in Fraser et al. (2014) have essentially no effect. Note that because we compare our results 
with the CFEPS survey, all absolute magnitudes reported here are considered to be in the g band. 
\section{Results}
All migration simulations were run to 0.5 Gyr. They were extended to 2 or 4 Gyr with the standard {\tt swift\_rmvs4} code (i.e., 
without migration/damping after 0.5 Gyr). We performed 16 new simulations in total. Four of these simulations
considered the case with smooth migration. In case 1, we used $\tau_1=30$ Myr and $\tau_2=100$ Myr. In case 2, we used $\tau_1=10$ 
Myr and $\tau_2=30$ Myr. For each of these cases, we performed 
a simulation with $\Delta a_{\rm N}=0$, and another simulation with $\Delta a_{\rm N}=0.5$ AU. In addition, we performed 12 simulations 
with the grainy migration. 
These simulations shared the properties of the four smooth migration cases, but for each case we considered
several different assumptions on the migration graininess. Specifically, the outer disk was assumed to have 1000, 2000 or 4000 
massive planetesimals each with a Pluto mass ($M_{\rm mp}=M_{\rm Pluto}$), or 1000 massive planetesimals each with twice the mass of 
Pluto ($M_{\rm mp}= 2 M_{\rm Pluto}$; hereafter Twopluto). A more detailed exploration of parameter space was not possible, because 
each simulation with $10^6$ disk particles is computationally expensive (one full simulation for 4 Gyr requires $\sim$5000 CPU 
days on NASA's Pleiades Supercomputer\footnote{http://www.nas.nasa.gov/hecc/resources/pleiades.html}).

For the population estimates discussed below we first need to define what we mean by different 
categories of KBOs. The HCs are defined here as objects on orbits with semimajor axis $40<a<47$ AU and perihelion distance 
$q=a(1-e)>36$ AU. Using a smaller perihelion distance cutoff would not affect the population estimate much, because there 
are not that many orbits with $q<36$ AU in the quoted semimajor axis range. We do not make any effort to separate the HCs from 
the populations in the 5:3, 7:4, 9:5, 11:6 resonances that intersect the main belt. This should not be a problem either, 
because the populations in these weak resonances are much smaller than the HC population.

As for the 3:2 resonance, we require that the resonant angle, $\sigma_{3:2}=3\lambda-2\lambda_{\rm N}-\varpi$, where $\lambda$ and 
$\varpi$ are the mean and perihelion longitudes of a particle, and $\lambda_{\rm N}$ is the mean longitude of Neptune, 
librates. This is done by selecting all particles with $38.5<a<40$~AU at the end of our simulations, performing an additional 
$10^6$-yr simulation for them, and computing the libration amplitude, $A_\sigma$, as half of the full range of the $\sigma_{3:2}$
excursions. The maximum amplitude of stable librations seen in our simulations is $A_\sigma\simeq130^\circ$, which is similar
to the maximum libration amplitudes of known Plutinos (Nesvorn\'y \& Roig 2001, Gladman et al. 2012). 
The non-librating orbits with $38.5<a<40$ AU typically have low eccentricities, because the low eccentricities are required 
near the 3:2 resonance for the orbital stability.  
%
%
\subsection{Resonance Overpopulation Problem}
\begin{figure}
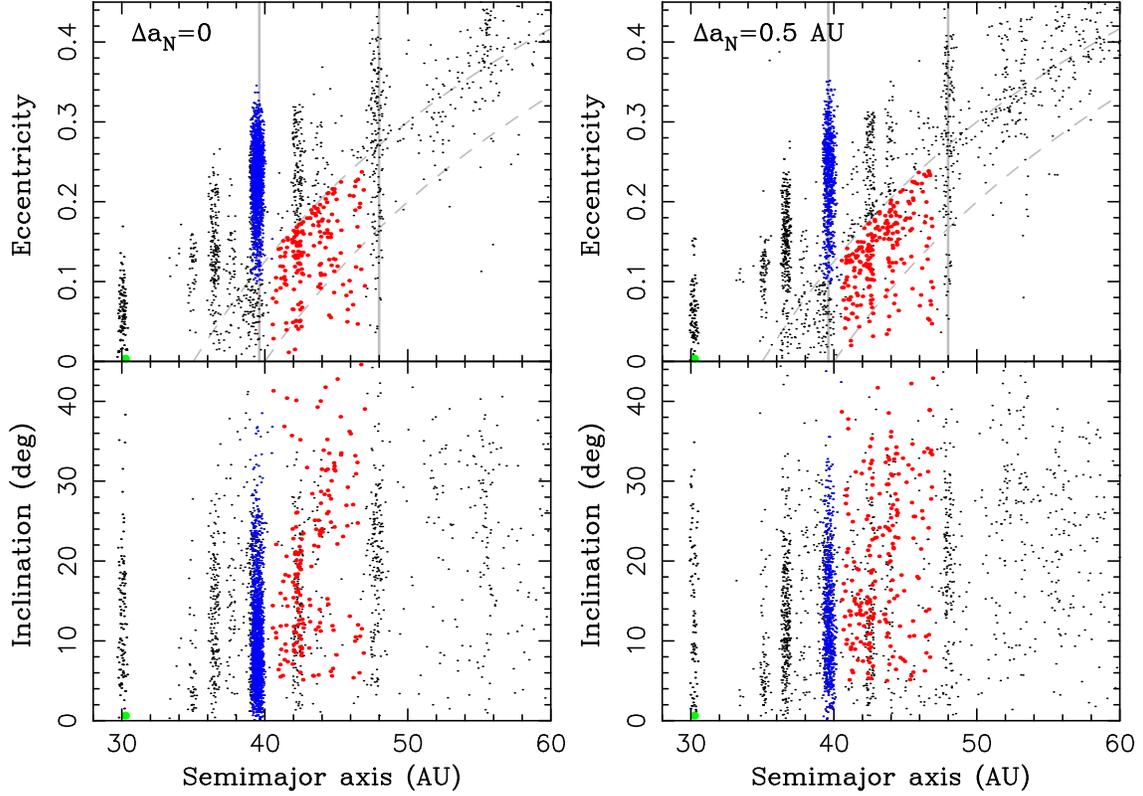

\epsscale{0.45}
\plotone{fig6a.eps}
\plotone{fig6b.eps}
\caption{The orbital elements of bodies captured in the Kuiper with smooth migration and the case-1 parameters 
($\tau_1=30$ Myr and $\tau_2=100$ Myr). The left panels show the result for $\Delta a_{\rm N}=0$ and the right panels 
show the result for $\Delta a_{\rm N}=0.5$~AU. Orbits in the 3:2 resonance and in the main belt ($40.5<a<47$ AU)
are highlighted by blue and red colors, respectively. The two vertical lines in the upper panels show the
positions of the 3:2 and 2:1 resonances with Neptune. Note that the resonances are strongly overpopulated relative 
to the HCs (cf. Fig. 1). There are roughly 7 (3) times as many Plutinos as the HCs for $\Delta a_{\rm N}=0$ 
($\Delta a_{\rm N}=0.5$~AU). This contradicts observations.}
\label{smooth2}
\end{figure}

We first discuss whether Neptune's jump, as suggested in Nesvorn\'y (2015a), can help to resolve the resonance overpopulation problem. 
Figure \ref{smooth2} shows the final distribution of orbits implanted into the Kuiper belt in case 1 with smooth migration. 
For $\Delta a_{\rm N} = 0$ (i.e., no jump) we obtain $N_{\rm HC}/N_{3:2}=0.14$, while for $\Delta a_{\rm N} = 0.5$ AU we find 
$N_{\rm HC}/N_{3:2}=0.35$ (Table 1). Thus, the ratio increased by a factor of 2.5 when Neptune's jump was accounted for 
in the model. The main difference between these two cases is that the probability of capture on a stable orbit in the 3:2 resonance, 
$P_{3:2}$, is $P_{3:2}=2.0\times 10^{-3}$ for $\Delta a_{\rm N} = 0$ and $P_{3:2}=6.8\times10^{-4}$ for $\Delta a_{\rm N} = 0.5$ AU.
[The probabilities are normalized here to one particle in the original disk below 30 AU.] 

We looked into this issue in detail and found that the change of $P_{3:2}$ was mainly contributed by Neptune's jump. As a consequence of 
Neptune's jump, the 3:2 resonant objects captured during the previous stage were released from the resonance. Interestingly, however, many 
bodies were captured into the 3:2 resonance from the scattered disk immediately after Neptune's jump, when the 3:2 resonance suddenly moved 
into a new orbital location, and during the subsequent slow migration of Neptune. This explains why the $N_{\rm HC}/N_{3:2}$ ratio did not 
change more substantially. The probability of capture in the main belt, $P_{\rm HC}$, also changed, but the change was minor 
($P_{\rm HC}=2.8\times 10^{-4}$ for $\Delta a_{\rm N} = 0$ and $P_{\rm HC}=2.4\times10^{-4}$ for $\Delta a_{\rm N} = 0.5$ AU; Table 1).

A similar result holds for case 2 with smooth migration, where $N_{\rm HC}/N_{3:2}=0.45$ for $\Delta a_{\rm N} = 0$ and $N_{\rm HC}/N_{3:2}=0.66$ 
for $\Delta a_{\rm N} = 0.5$ AU. These values are somewhat higher that those obtained in case 1 perhaps suggesting that $N_{\rm HC}/N_{3:2}$ 
might increase further if even shorter migration timescales were used. The short migration timescales are not plausible, however, 
because they do not satisfy the inclination constraint (Nesvorn\'y 2015a). Given these results, we conclude that the effect of Neptune's 
jump can help, but it is insufficient in itself to resolve the resonance overpopulation problem. This is because even in the 
smooth migration cases with $\Delta a_{\rm N} = 0.5$ AU, the 3:2 resonance is still strongly overpopulated, by a factor of $\sim$5-10 
relative to HCs. Other resonances, such as the 2:1 or 5:2, are overpopulated by a significant factor as well. We therefore proceed by considering 
the cases with grainy migration.

\begin{figure}
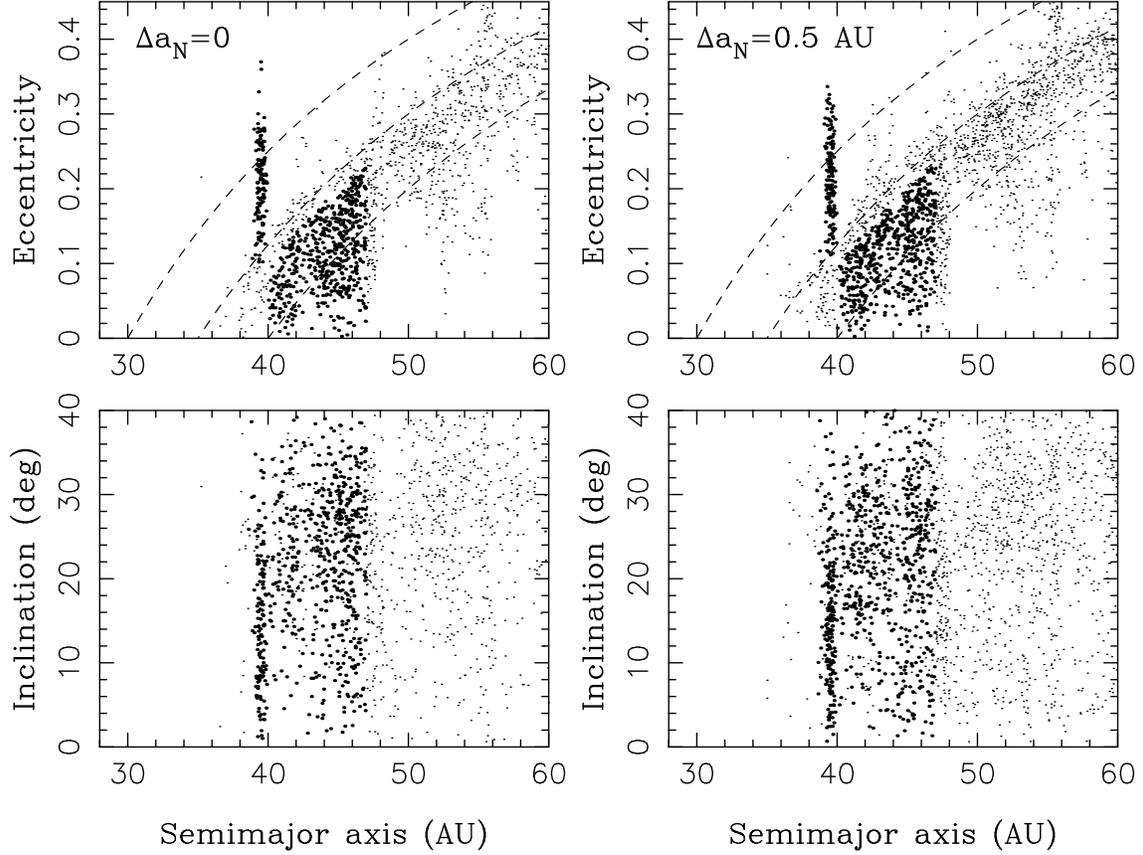

\epsscale{0.45}
\plotone{fig7a.eps}
\plotone{fig7b.eps}
\caption{The orbital elements of bodies captured in the Kuiper belt in a model with the case-1 migration timescales
($\tau_1=30$ Myr and $\tau_2=100$ Myr), grainy migration corresponding to 1000 massive planetesimals each with mass 
$M_{\rm mp}=2\ M_{\rm Pluto}$, $\Delta a_{\rm N}=0$ (left panels) and $\Delta a_{\rm N}=0.5$~AU (right panels). The HCs 
and Plutinos are denoted by larger symbols.}
\label{grainy}
\end{figure}

Figure \ref{grainy} shows some of our best results for the grainy migration. These results were obtained for case 1 ($\tau_1=30$ Myr and 
$\tau_2=100$~Myr) and 1000 Twoplutos. Here, $N_{\rm HC}/N_{3:2} \simeq 4$ for both $\Delta a_{\rm N} = 0$ and $\Delta a_{\rm N} = 0.5$ AU, in a 
close match to CFEPS observations. This is encouraging. Neptune's jump does not seem to have much to do with this result, because 
$N_{\rm HC}/N_{3:2} \simeq 4.2$ for $\Delta a_{\rm N} = 0$ and $N_{\rm HC}/N_{3:2} \simeq 3.8$ for $\Delta a_{\rm N} = 0.5$~AU. These two values 
are similar and the small difference between them probably reflects some minor difference in the planetary migration histories. Neptune's jump 
has only a small effect here, because the case with grainy migration encourages late captures with many bodies being captured {\it after} the 
instability. 

The results for the case-1 grainy migration with 4000 Plutos are similar to those reported above for 1000 Twoplutos. The ones obtained 
for 1000 and 2000 Plutos are intermediate showing a clear progression from the smooth migration case to cases with an increased migration 
graininess (Table 1). The main trend is that $P_{3:2}$ drops when more and more Plutos are included. 
For case 1 with 1000 Plutos, $P_{3:2}=8.9 \times 10^{-4}$ for $\Delta a_{\rm N}=0$ and $P_{3:2}=3.2\times10^{-4}$ for $\Delta a_{\rm N}=0.5$~AU. 
With 2000 or 4000 Plutos, or 1000 Twoplutos, $P_{\rm 3:2} \simeq (1.5$-$4)\times 10^{-4}$, which represents only a small fraction of 
$P_{\rm 3:2}$ obtained with the smooth migration. This change is attributed to the jittery motion of the 3:2 resonance that 
accompanies Neptune's grainy migration (Murray-Clay \& Chiang 2006). Due to this jitter, many bodies captured into the resonance 
during the earlier stages were later released from the resonance, because their libration amplitude increased beyond the 
limits required for the orbital stability ($A_\sigma\simeq130^\circ$; Nesvorn\'y \& Roig 2001). The distribution of the libration amplitudes in the 
3:2 resonance is discussed in Section 3.3.   

The HC capture probability also changes when the migration graininess is increased in the model. With case 1 and 1000 Plutos, 
$P_{\rm HC}\simeq4$-$5\times10^{-4}$, about a factor of 2 higher than in the smooth migration case. The capture probability 
increases further, to $P_{\rm HC}\simeq 6\times10^{-4}$, when 2000 or 4000 Plutos are included. Together, the increasing 
$P_{\rm HC}$  and decreasing $P_{3:2}$ lead to $N_{\rm HC}/N_{3:2}$ values that are more in line with observations. The best fit 
to observations, with $N_{\rm HC}/N_{3:2} \simeq 2$-4, occurs for the case-1 simulations with 2000 and 4000 Plutos (Table~1).
It is difficult to infer the precise number of Pluto-mass bodies with more confidence. On one hand, additional observations 
are needed to better constrain the $N_{\rm HC}/N_{3:2}$ ratio in the present Kuiper belt. On the other hand, the massive planetesimals 
in the original disk must have had a range of masses, while here we represent their size distribution by a delta function.
A more realistic modeling with a continuous mass distribution of massive planetesimals is left for future work. See Gladman 
\& Chan (2006) for a modeling work of the effect of very massive planetesimals.  

For an outer disk with mass $M_{\rm disk}=15$-20 $M_{\rm Earth}$, $P_{\rm HC} \simeq 6\times10^{-4}$ in our reference case with grainy migration implies 
the HC mass $M_{\rm HC}=0.008$-0.012 $M_{\rm Earth}$, while the mass inferred from observations is $M_{\rm HC} \simeq 0.01$ $M_{\rm Earth}$ (Fraser 
et al. 2014). This is an excellent agreement. Since the model population of Plutinos in the 3:2 resonance has the right proportion relative to 
the HCs, as discussed above, this implies that the Plutino mass obtained in the simulation is also approximately correct. 

\begin{figure}
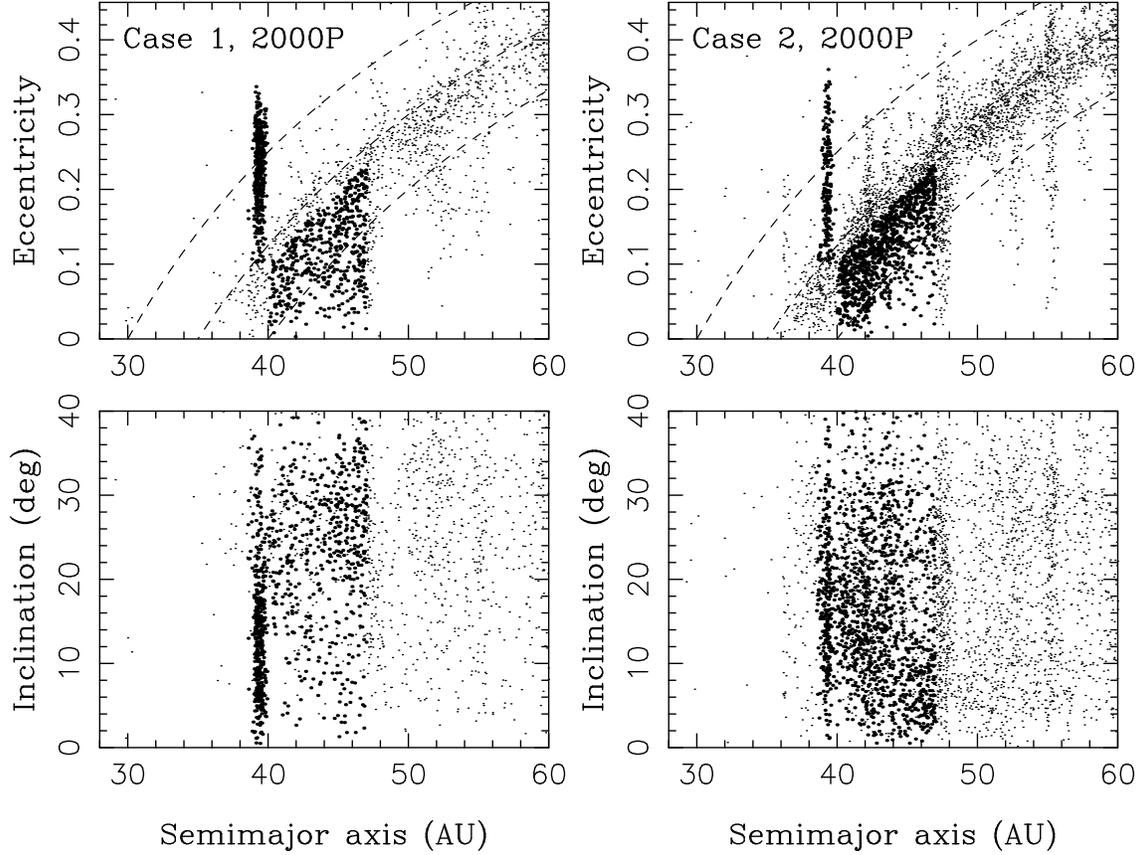

\epsscale{0.45}
\plotone{fig8a}
\plotone{fig8b.eps}
\caption{The orbital elements of bodies captured in the Kuiper belt in a model with grainy migration corresponding to 
2000 massive planetesimals each with mass $M_{\rm mp}=M_{\rm Pluto}$, and $\Delta a_{\rm N}=0.5$~AU. The left and right panels show the 
results for case 1 ($\tau_1=30$ Myr and $\tau_2=100$~Myr) and case 2 ($\tau_1=10$ Myr and $\tau_2=30$~Myr), respectively.
The HCs and Plutinos are denoted by larger symbols.}
\label{2000}
\end{figure}

The case 2 with $\tau_1=10$ Myr and $\tau_2=30$~Myr shows trends in many ways similar to those discussed for case 1 above. 
The principal effect of faster migration rates is to produce the $P_{\rm HC}$ values that are a factor of $\sim$2 larger than in 
case 1 (for the same level of graininess), and $P_{3:2}$ values that are somewhat smaller. Together, these trends make it easier
to obtain the observed $N_{\rm HC}/N_{3:2} \simeq 2$-4 with a lower level of graininess. This is illustrated in Fig. \ref{2000}, where we 
show the orbital distributions obtained with 2000 Plutos. We find that $N_{\rm HC}/N_{3:2}\simeq1.5$ in case 1 and $N_{\rm HC}/N_{3:2}\simeq6$
in case 2. Thus, while the case 1 with 2000 Plutos produces a ratio that is slightly lower than the one indicated by observations, the 
case 2 with 2000 Plutos overshoots it. The best results in case 2 were obtained with 1000 Plutos in the original disk. In this case,
$N_{\rm HC}/N_{3:2}=1.8$ for $\Delta a_{\rm N}=0$ and $N_{\rm HC}/N_{3:2}=2.5$ for $\Delta a_{\rm N}=0.5$ AU (Table 1).

We conclude that the statistics inferred from observations of the resonant and non-resonant populations in the Kuiper
belt implies that the massive planetesimal disk below 30~AU contained 1000-4000 Plutos. The combined probability that a planetesimal 
from the original disk below 30 AU evolves on a Kuiper belt orbit is $\sim$$10^{-3}$. With 1000-4000 Plutos in the original
disk, we would therefore expect that $\sim$1-4 Pluto-class objects should exist in the Kuiper belt today, while two 
such objects are known (Pluto and Eris). This is a reasonable agreement, but note that neither Pluto or Eris is a member
of the HC population, while we would expect from our model that the Pluto-size objects are preferentially deposited in the HC 
population. The expectations would be slightly different if a 
continuous mass distribution of massive planetesimals were considered. For example, it is plausible that the needed migration graininess 
was produced by the combined effect of 1000 Plutos {\it and} 500 Twoplutos. This would yield $\sim$1 Pluto and $\sim$0.5 Twoplutos in 
the Kuiper belt today. Obviously, these considerations are subject to the small number statistics. Their main point is to show that the 
needed graininess from the $N_{\rm HC}/N_{3:2}$ constraint is not contradictory to having two Pluto-class objects in the Kuiper 
belt today.     
\subsection{The Inclination Distribution}
The implantation of the disk planetesimals into the Kuiper belt is a multi-step dynamical process that was first pointed 
out in Gomes (2003), and is hereafter called the Gomes mechanism. In the Gomes mechanism, disk planetesimals are first 
scattered by Neptune to $>$30 AU, where they can evolve onto orbits with large libration amplitudes in mean 
motion resonances. The secular dynamics inside the mean motion resonances, including the Kozai cycles (Kozai 1962),
can subsequently act to raise the perihelion distance and decouple the orbits from Neptune. Finally, if Neptune is still 
migrating, bodies can be released from the resonances onto stable non-resonant orbits. While the Gomes mechanism can 
operate for a wide range of migration parameters, Nesvorn\'y (2015a) found that the inclination constrain requires 
that Neptune is given sufficient time to act on the scattered bodies and increase their orbital inclinations {\it before}
bodies decouple from Neptune (the Kozai cycles inside mean motion resonances also contribute to increasing the 
orbital inclinations, but they are not the principal factor). Hence it is required that Neptune's migration was slow.

Nesvorn\'y (2015a) used a slightly different migration setup from the one utilized here. The migration in their case was 
smooth (i.e., no massive bodies in the disk) and characterized by a single migration phase with the starting position of 
Neptune $a_{\rm N,0}$  and e-folding migration timescale $\tau$. They found that the inclination constraint implies that 
$a_{\rm N,0} \lesssim 25$ AU and $\tau \gtrsim 10$ Myr. 
The migration recipe used in this work was described in Section 2. Here we have two migration stages with a slow migration 
during the first stage, and even slower migration during the second stage. The migration timescales used for the two 
phases satisfy the inclination constraint because $\tau_{1,2} \geq 10$ Myr. At the end of the first phase, we assumed that 
the orbit of Neptune may have changed discontinuously. And, in addition, our preferred migration mode is grainy.
These differences could affect the inclination distribution. Here we therefore test whether the orbital distribution 
obtained with our favored migration parameters satisfies the inclination constraint.   

\begin{figure}
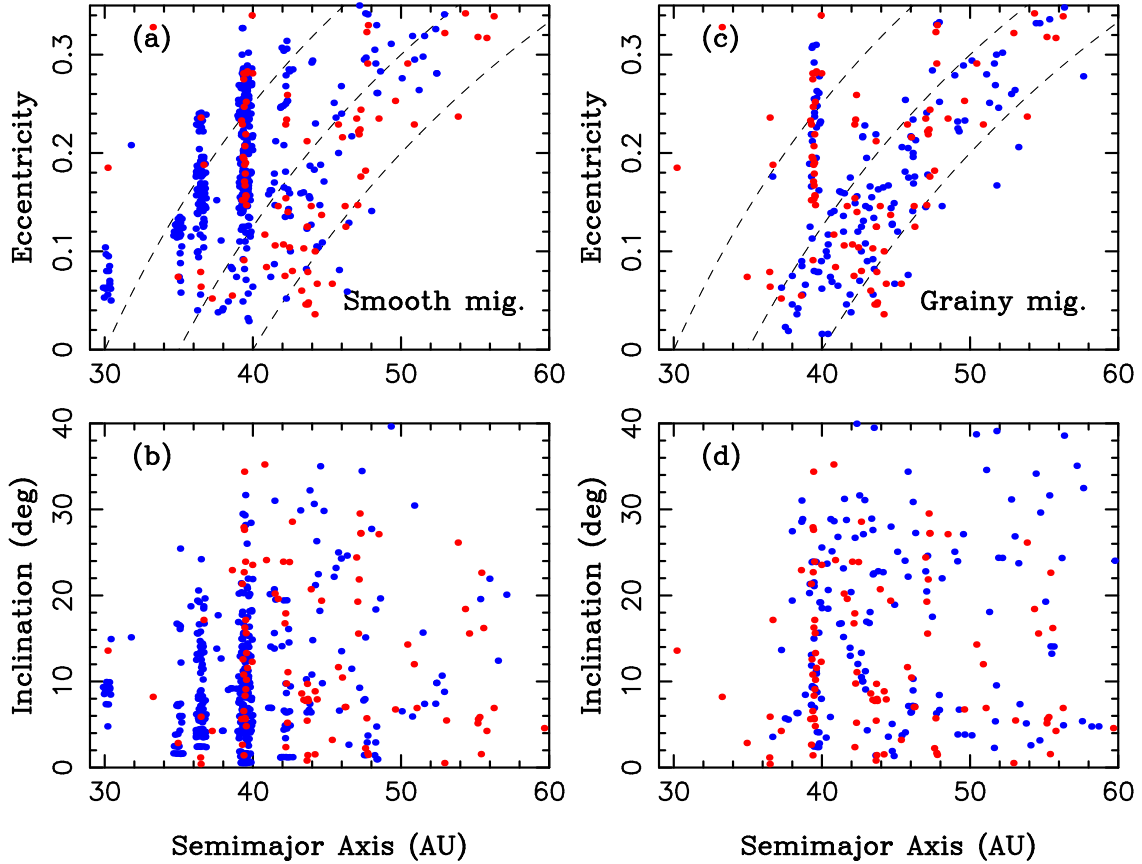

\epsscale{0.45}
\plotone{fig9a.eps}
\plotone{fig9b.eps}
\caption{A comparison of the orbital distributions obtained in our model (blue dots) and the actual CFEPS 
detections (red dots). The left panels show the distribution obtained for a smooth migration of Neptune. The right 
panels show the result obtained with grainy migration assuming that there were 1000 massive
planetesimals with $M_{\rm mp}=2\ M_{\rm Pluto}$ in the original disk. In both cases, we used $\tau_1=30$ Myr, 
$\tau_2=100$ Myr, $a_{\rm N,1}=27.8$ AU, $\Delta a_{\rm N} =0.5$~AU, $\Delta e_{\rm N} = 0.1$. The CFEPS 
detection simulator was applied to the model, and the resulting distribution of the detected orbits is shown 
here. For the actual CFEPS detections, we plot all orbits that were {\it not} classified by the CFEPS team 
as belonging to the CC population.}
\label{cfeps}
\end{figure}

Figure \ref{cfeps} illustrates how the orbital distribution of bodies obtained in our case-1 simulation ($\tau_1=30$ Myr,
$\tau_2=100$ Myr, $\Delta a_{\rm N}=0.5$ AU) compares with observations. Two cases are shown: (1) a grainy migration case 
with 1000 Twoplutos, and (2) a smooth migration case for a reference. We used the CFEPS detection simulator, as described in Section 2.4, 
and compared the simulated orbits with the actual CFEPS detections. The agreement is satisfactory in the grainy case,
where the distribution of the non-resonant orbits roughly follows the lines of constant perihelion distance such that a larger 
value of the semimajor axis implies larger eccentricity. This trend is a characteristic property of the HC population
(see Nesvorn\'y 2015a for a more detailed comparison of our model with the CFEPS observations).
The 3:2 resonance population obtained in the grainy migration case has a correct distribution of orbital eccentricities. Moreover, 
as we discussed in Section 3.1, the HCs and resonant populations appear in the right proportion. For comparison, the 
smooth migration case shown in Fig.~\ref{cfeps}a,b leads to an excessive number of detections in resonances, which 
is clearly incorrect.  

\begin{figure}
\epsscale{0.45}
\plotone{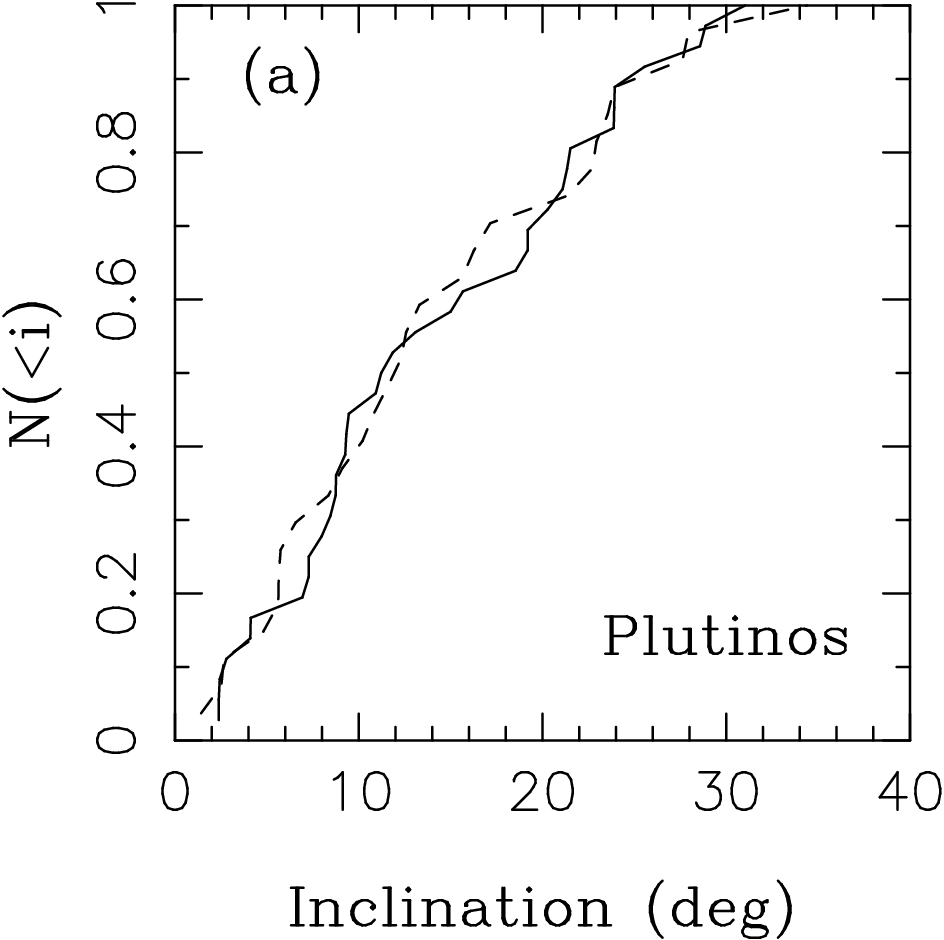}
\plotone{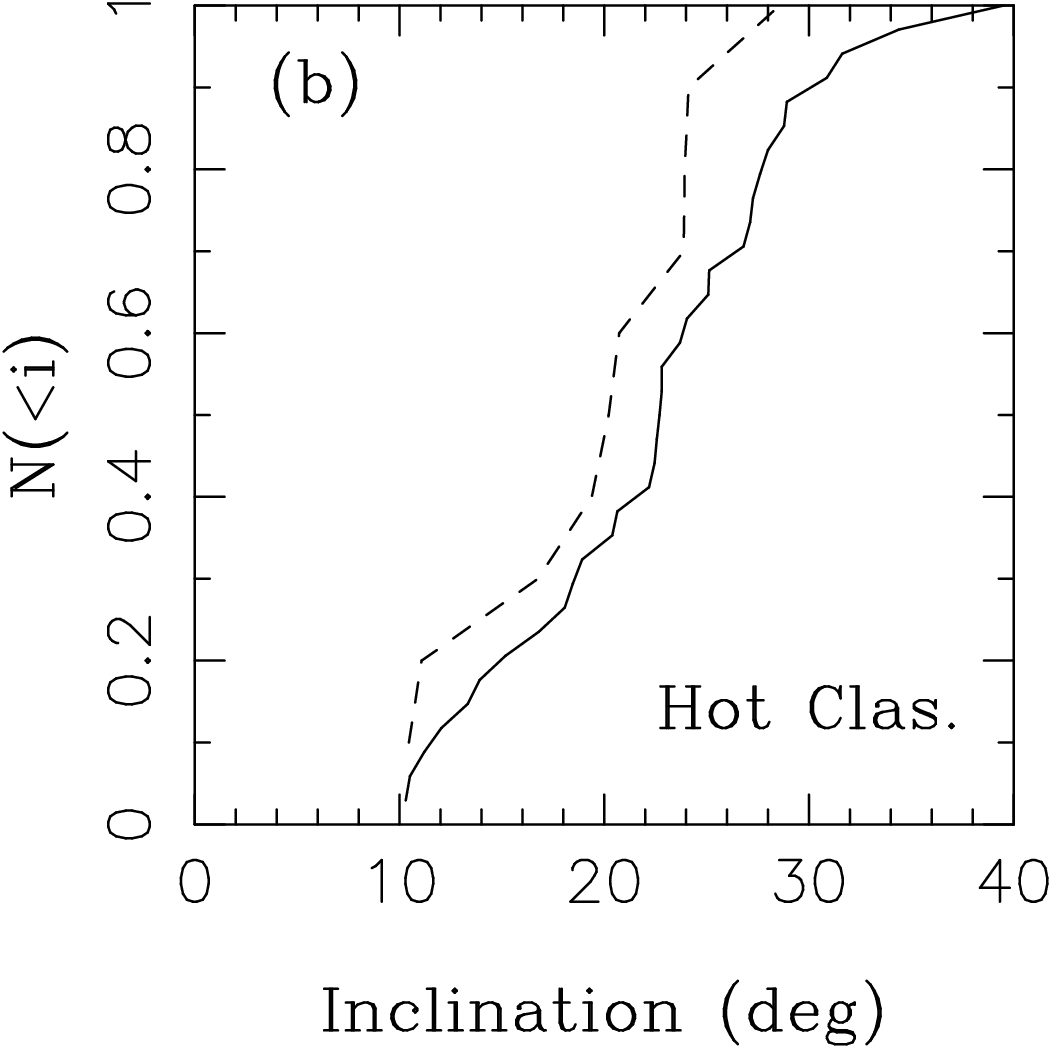}
\caption{A comparison of the inclination distributions obtained in our model (solid lines) and the CFEPS 
detections (dashed lines). Here we used $\tau_1=30$ Myr, $\tau_2=100$ Myr, $a_{\rm N,1}=27.8$ AU, $\Delta 
a_{\rm N} =0.5$ AU, $\Delta e_{\rm N} = 0.1$, and the migration graininess corresponding to 1000 massive
planetesimals each with $M_{\rm mp}=2\ M_{\rm Pluto}$. The CFEPS detection simulator was applied to the model 
orbits to have a one-to-one comparison with the actual CFEPS detections. In panel (b), we plot orbits with 
$i>10^\circ$ to avoid any potential contamination from the CCs.}
\label{idist}
\end{figure}

The inclination distribution obtained in the model is wide and roughly comparable to the one inferred from observations.
A more careful comparison of the inclination distributions is presented in Fig. \ref{idist}, where the model distributions 
are shown to follow very closely the CFEPS distributions. This is especially true for the inclination distribution of 
Plutinos (Fig. \ref{idist}a), for which the K-S test gives a 84\% probability that the simulated and observed 
distributions are being derived from the same underlying distribution. The agreement is somewhat less satisfactory for 
HCs, where the model distribution is slightly wider than the observed one and the K-S test gives a 23\% probability.
Still, this is a satisfactory match. Also, note that the inclination distribution is sensitive to the migration timescale, 
and slightly shorter migration timescales should lead to a better agreement (Nesvorn\'y 2015a). 

Indeed, our case 2 with $\tau_1=10$ Myr and $\tau_2=30$ Myr yields a narrower inclination distributions of the HCs 
(Fig. \ref{idist2}b). In this case, the K-S test gives the 57\% probability for Plutinos and 80\% probability for
the HCs. The main difference with respect to case 1 is that the model distribution of HCs now represents a reasonable match 
to observations all the way down to $i\simeq5^\circ$, while in case 1 we were able to produce a satifactory fit only
for $i>10^\circ$ (Nesvorn\'y 2015a). This may indicate that the real migration timescales were closer to case~2
than to case 1.

In case 1 with 2000 and 4000 Plutos the inclination distributions are similar to the one shown in Fig. \ref{idist}. 
All other cases studied here show narrower inclination distributions for Plutinos. Cases 1 and 2 with a smooth migration 
also show narrower inclination distributions of HCs, which nicely fit the observed distribution for $5^\circ < i < 10^\circ$, 
where the CFEPS inclination distribution steeply raises, but fail to match the wide distribution for $i>10^\circ$. This shows 
that there is some trade-off between the level of graininess and the migration timescale. Any future attempt to closely 
match the inclination distribution will thus need to explore both these parameters with more resolution. Here we content 
ourselves with showing that the general results published in Nesvorn\'y (2015a) are valid even if Neptune's migration 
was grainy.           
\subsection{Distribution of Libration Amplitudes}
The distribution of libration amplitudes in the 3:2 resonance was characterized by the CFEPS (Gladman et al. 2012).
According to Fig. 3 in Gladman et al. (2012), the cumulative distribution of libration amplitudes, $A_\sigma$, appears to be steadily 
raising from $A_\sigma \simeq 20^\circ$ to $A_\sigma \simeq 100^\circ$, and tails off for $A_\sigma > 100^\circ$. 
In terms of a differential distribution, Gladman et al. (2012) suggested an asymmetric triangle model, where 
the number of orbits in a $\Delta A_\sigma$ interval linearly increases from zero at $A_\sigma = 20^\circ$ to a maximum 
at $A_\sigma \simeq 90^\circ$, and then linearly drops to zero at $A_\sigma \simeq 130^\circ$, because the orbits with 
$A_\sigma > 130^\circ$ are unstable.

Gladman et al. (2012) also pointed out that the distribution of $A_\sigma$ inferred from the CFEPS observations is very similar 
to the theoretical distribution reported in Nesvorn\'y \& Roig (2001), where the 3:2 resonance was randomly populated, and 
the orbital distribution was dynamically evolved over 4 Gyr (with Neptune on its current orbit). The number 
of resonant orbits increases with $A_\sigma$, because the resonant orbits with larger libration amplitudes 
represent a larger phase-space volume than the orbits with smaller libration amplitudes, and are therefore more populated 
to start with. There are fewer orbits with $A_\sigma > 100^\circ$, because this is already close to the stability limit, and 
the original population was depleted when the orbits with $A_\sigma > 100^\circ$ evolved out of the resonance.  
  
Figure \ref{sigma} shows the distributions of libration amplitudes in the 3:2 resonance obtained in the models with smooth 
and grainy migrations. The grainy migration case matches the observed distribution much better that the smooth case.
The K-S probabilities obtained for the grainy case are 32\% for $\Delta a = 0$ and 17\% for $\Delta a = 0.5$ AU,
while the probability in the smooth case is $\sim$$10^{-3}$. This result, in itself, could be used rule out the smooth migration 
case, where the libration amplitudes are significantly larger than the ones found by the CFEPS (Gladman et al. 2012). This is a 
consequence of the capture mechanism in the 3:2 resonance, which tends to produce large amplitudes if the migration is 
smooth. The amplitudes in the grainy case are a bit larger than what would be ideal for $\Delta a = 0.5$ AU, and the 
distribution is somewhat shallower for $\Delta a = 0$, but we do not consider these slight differences being 
significant.\footnote{Note that the distribution of the libration amplitudes can be modified over very long time 
scales by the gravitational encounters of Plutinos with Pluto (e.g., Yu \& Tremaine 1999, Nesvorn\'y et al. 2000). 
Here we ignore this effect.}

The distributions of $A_\sigma$ obtained for the smooth and grainy migration cases are significantly different in all cases studied 
here. The amplitude distribution obtained for the smooth-migration case 2 is similar to the one shown in Fig. \ref{sigma} for 
the smooth-migration case 1 (blue line). This is independent of whether or not Neptune jumped during the instability. 
The smooth migration cases therefore produce, in general, the amplitude distributions that do not agree with observations.
With the grainy migration corresponding to 4000 Plutos, on the other hand, the model distributions of $A_\sigma$ become slightly 
narrower than what is inferred from observations. From this we conclude that 1000-4000 Plutos give the best fit to 
observations. It is encouraging to see that while this argument is independent of the one based on the population statistics 
(Section 3.1), it leads to a similar inference about the number of Pluto-sized objects. In any case, these results represent 
a significant improvement over those shown in Fig. 7 of Gladman et al. (2008), where the amplitude distribution obtained in 
the model of Levison et al. (2008; $a_{\rm N,0}=28$ AU, smooth migration  with $\tau=1$ Myr) was shown to be strongly discordant 
with the CFEPS observations.
\subsection{The Cold Classicals and Kernel}
The CCs have low orbital inclinations ($i<5^\circ$) and several 
physical properties (red colors, large binary fraction, steep size distribution of large objects,
relatively high albedos) that distinguish them from all other KBO populations.\footnote{Specifically, (1) the CCs have 
distinctly red colors (e.g., Tegler \& Romanishin 2000) that may have resulted from space weathering of surface ices, such 
as ammonia (Brown et al. 2011), that are stable beyond $\sim$35 AU. (2) A large fraction of the 100-km-class CCs are wide 
binaries with nearly equal size components (Noll et al. 2008). (3) The albedos of the CCs are generally higher 
than those of the HCs (Brucker et al. 2009). And finally, (4) the size distribution of the CCs is markedly different 
from those of the hot and scattered populations, in that it shows a very steep slope at large sizes (e.g., 
Bernstein et al. 2004, Fraser et al. 2014), and lacks very large objects (Levison \& Stern 2001).}
The most straightforward interpretation of the unique physical and orbital properties is that the CCs formed and/or 
dynamically evolved by different processes than other trans-Neptunian populations. Here we consider a possibility that 
the CCs formed at $>$40~AU and survived Neptune's early `wild days' relatively unharmed (e.g., Kavelaars et al. 2009, 
Batygin et al. 2011, Wolff et al. 2012). This requires that the massive planetesimal disk at $<$30 AU had a low-mass 
extension beyond 30~AU, as already discussed in Section 2.2. Nesvorn\'y (2015b) studied this model and found that the original disk at 42-47 
AU only contained the mass $\sim$$6 \times 10^{-3}$ $M_{\rm Earth}$. The surface density of solids in this region, $\Sigma_{\rm s} 
\sim 2 \times 10^{-5}$ g cm$^{-2}$, was probably therefore some $\sim$3000 times lower than in the massive part of the disk 
below 30 AU. This implies that the CCs must have formed by an efficient accretion mechanism that was capable of building
$\sim$100-km planetesimals in a low-mass environment  (e.g., Johansen et al. 2009).\footnote{The polution of CCs from the HCs 
should be minimal, because only $\sim$4\% of the HC orbits obtained in our model have $i<5^\circ$. This represents $\sim$0.0004 
$M_{\rm Earth}$, or only about 10\% of the estimated mass of the CC population ($\sim$0.003 $M_{\rm Earth}$ according to 
Fraser et al. 2014).} 

According to Petit et al. (2011), the CC population can be divided into the `stirred' and `kernel' 
components. The stirred orbits have the semimajor axes $42.4<a<47$ AU, inclinations $i<5^\circ$, and 
small eccentricities with an upper limit that raises from $e\simeq0.05$ for $a=42$ AU to $e\simeq0.2$ for $a=47$ AU.
The kernel is a narrow concentration of low-inclination orbits with $a\simeq44$ AU, $e\simeq0.05$, and 
a $\simeq$0.5-1 AU width in the semimajor axis. 
Figure \ref{kernel} illustrates a model of the orbital distribution inferred from the CFEPS observations.  

Nesvorn\'y (2015b) suggested that the Kuiper belt kernel can be explained if Neptune's 
otherwise smooth migration was interrupted by a discontinuous change of Neptune's semimajor axis when Neptune reached 
$\simeq$28~AU. Before the discontinuity happened, planetesimals located at $\sim$40 AU were swept into Neptune's 
2:1 resonance, and were carried with the migrating resonance outwards. The 2:1 resonance was at $\simeq$44 AU 
when Neptune reached $\simeq$28~AU. If Neptune's semimajor axis changed by fraction of AU at this point, 
perhaps because Neptune was scattered off of another planet (see Fig. 3), the 2:1 population would have been released 
at $\simeq$44 AU, and would remain there to this day. The orbital distribution produced in this model provides 
a good match to the orbital properties of the kernel. 
 
Nesvorn\'y (2015b) model assumptions and migration parameters were the same as in this work, except that (1) they 
considered a low mass extension of the planetesimal disk at 30-50 AU, and (2) their migration was ideally smooth, while 
here we showed that the population statistics inferred from observations requires that the migration was grainy. We 
therefore repeat the simulations of Nesvorn\'y (2015b) to test whether the kernel can form even if the migration 
was grainy.

Each simulation included 5000 test particles distributed from 30 to 50 AU. Their radial profile was set such that the disk 
surface density $\Sigma \propto 1/r$. There is therefore an equal number of particles (250) in each radial AU. A larger 
resolution is not needed, because a significant fraction of particles in the CC region survive. The disk extension was 
assumed to be dynamically cold with low orbital eccentricities and low orbital inclinations. The initial inclinations 
were set to be similar to those inferred for the present population of CCs. Specifically, 
we used $N(i)\,{\rm d}i = \sin i \exp(-i^2/2\sigma_i^2)\,{\rm d}i$, with $\sigma_i=2^\circ$ (Brown 2001, Gulbis et al. 2010). 
The initial eccentricities were set according to the Rayleigh distribution with $\sigma_e=0.01$, 0.05 or 0.1. 

Figure \ref{colds} shows the orbital distribution of particles obtained in a model with $\sigma_e=0.01$, and the case-1 
migration parameters with the graininess corresponding to 1000 Twoplutos. The result is similar to those published in 
Nesvorn\'y (2015b) for the smooth migration. The concentration of orbits near 44 AU has the orbital properties comparable 
to those of the CFEPS kernel. This shows that the grainy migration required to explain the population statistics of resonant and 
non-resonant populations also allows for the formation of the Kuiper belt kernel. These results are therefore consistent with
each other. The concentration of orbits obtained in the model near 44 AU becomes slightly more fuzzy for $\sigma_e=0.05$ or 
0.1, following the trends described in Nesvorn\'y (2015b). The case-2 parameters with grainy migration also lead to the 
formation of the kernel. We therefore conclude that the model of kernel formation described in Nesvorn\'y (2015b) does not 
require that the migration was smooth. Instead, it works even if the migration was grainy.   
\subsection{The 2:1 and 5:2 Resonances}
Adams et al. (2014) found from the Deep Ecliptic Survey (DES) that $N_{3:2}/N_{2:1} \simeq N_{3:2}/N_{5:2} \simeq 2$, while 
Gladman et al. (2012) suggested from the CFEPS that $N_{3:2}/N_{2:1} \simeq 3$-4 and 
$N_{3:2}/N_{5:2} \simeq 1$. Part of these differences between the DES and CFEPS may stem from differences in observational 
strategies and/or debiasing approach. While there is obviously a significant uncertainty in these estimates, it is probably 
fair to say that observations suggest roughly comparable populations in the 2:1 and 5:2 resonances (to within a factor of 
$\sim$2 or so), both of which are $\simeq$1-4 times smaller than Plutinos in the 3:2 resonance.  

Here we find that the smooth migration cases produce $N_{3:2}/N_{2:1}\sim 15$ and $N_{3:2}/N_{5:2}\sim 10$. The 2:1 and 5:2 
resonances are therefore clearly underpopulated, relative to the 3:2 resonance, if the migration is assumed to be smooth.
Much better results were obtained for the grainy migration. In case 2 and 2000 Plutos,  where the lowest resonant ratios were 
found, $N_{3:2}/N_{2:1}\simeq2$-2.5 and $N_{3:2}/N_{5:2}\simeq 1$. It is therefore plausible that the population of bodies
in the 5:2 resonance can be as large as Plutinos. Other results obtained here are intermediate. For example, case 1
gives $N_{3:2}/N_{2:1}\sim 10$ and $N_{3:2}/N_{5:2}\simeq6$ for 2000 Plutos, and $N_{3:2}/N_{2:1}\simeq3$-5 and $N_{3:2}/N_{5:2}\simeq2$ 
for 4000 Plutos. 

The increased level of migration graininess therefore leads to lower values of $N_{3:2}/N_{2:1}$ and
$N_{3:2}/N_{5:2}$, which are in better agreement with observations (Gladman et al. 2012, 2014; Adams et al. 2014). This trend
is mainly contributed by the lower value of $P_{3:2}$ when the migration is assumed to be grainy. On the other hand, 
if different migration timescales are considered for the same level of graininess, then the cases with faster migration 
rates tend to produce larger populations in the 2:1 ($a\simeq47.8$ AU) and 5:2 ($a\simeq55.5$ AU) resonances than the cases with slower migration rates.
This trend is clearly visible in Fig. \ref{2000}, where the 2:1 and 5:2 resonances are more populated in case 2 ($\tau_1=10$ Myr 
and $\tau_2=30$ Myr) than in case 1 ($\tau_1=10$ Myr and $\tau_2=30$ Myr).  
\section{Discussion}
In an attempt to develop a consistent model of Neptune's migration, we previously proposed that the wide inclination 
distribution of orbits inferred from observations can be explained if Neptune started inward of $\simeq25$ AU, and slowly 
(e-folding timescale $\tau \gtrsim 10$~Myr) migrated into a massive disk with the outer edge at $\simeq$30 AU. Moreover, 
we suggested that the concentration of low-inclination orbits at $\simeq$44 AU, known as the Kuiper belt kernel, can be 
explained if Neptune's semimajor axis discontinuously changed by $\simeq$0.5 AU when Neptune reached $a \simeq 28$ AU, 
perhaps because Neptune was scattered off of another planet (NM12). Here we pointed out that all previous models of 
the Kuiper belt formation suffered from the resonance overpopulation problem, where the resonant populations were overpopulated 
when compared to observations. We showed that this problem can be resolved if Neptune's migration was grainy as a result 
of close encounters of Neptune with massive, Pluto-class planetesimals. 

Here we considered the Pluto-class planetesimals because we have direct observational evidence that planetesimals such as 
Pluto or Eris exist in the present Kuiper belt. It is possible that Neptune's grainy migration was contributed by objects
much more massive than Pluto/Eris. We were not strongly motivated to consider, for example, an Earth-mass object in this work
(Gladman \& Chan 2006), because the overpopulation problem can be resolved by considering a reasonable number of smaller 
mass bodies (Plutos or Twoplutos). We thus really do not need to invoke effects of very massive planetesimals or planets. 
This does not exclude the possibility that such massive objects formed in the original disk, affected the 
dynamical evolution of Neptune, and helped to shape the orbital structure of the Kuiper belt (e.g., Gladman \& Chan 2006).
More detailed investigations of a continuous distribution of massive planetesimals, including cases with the Earth-class 
bodies, is left for future work. 

The best results were obtained if the massive disk below 30 AU was assumed to have contained 1000-4000 Plutos, or 
$\sim$1000 bodies twice as massive as Pluto. The total mass in these massive objects should thus be $\sim$2-8 $M_{\rm Earth}$, 
while the most plausible total mass of the disk was found to be $\simeq$20 $M_{\rm Earth}$ in NM12. This means that the 
Pluto-class objects should have represented $\simeq$10-40\% of the original disk mass. The remaining $\simeq$60-90\% 
of the mass was predominantly in the 100-km-class bodies, as inferred from the size distribution of the present 
Kuiper belt (e.g., Bernstein et al. 2004). To obtain this mass partitioning, a relatively steep size distribution of the 
planetesimal disk inferred from observations for diameters $\simeq$100-500 km cannot continue for $D>500$ km, because 
in that case the total mass in the $D>500$-km bodies would be negligible. Instead, the distribution needs to bulge at 
large sizes. 

Figure \ref{sfd} shows a reconstructed size distribution of the planetesimal disk below $\simeq$30~AU. We used several 
constraints here. For the intermediate sizes $10<D<500$~km, we adopted the size distribution suggested by Fraser et al. 
(2014) from observations of the Kuiper belt and Jupiter Trojans. This size distribution can be approximated by two
power laws with a break at $D\simeq100$ km, a steeper slope for larger sizes (cummulative power index $\simeq4.5$-5.0),
and a shallower slope for smaller sizes (cumulative index $\simeq1$-2). To estimate how the size distribution may have looked 
like for $D>500$ km, we assumed that there were 1000-4000 Plutos in the original disk, as required from the results 
of this study, and connected the size distribution from $D<500$ km to Pluto's diameter ($D=2370$ km). Note that the shallow
slope of the SFD in this range is consistent with observations of large KBOs (Brown 2008). The number of objects with 
$D>2500$ km drops in Fig. \ref{sfd}, but this part of the size distribution is unconstrained. The size distribution 
may have been shallower in this size range including some very massive objects in the original disk (e.g., Gladman \& Chan 2006). 

Only a few constraints exist for $D<10$ km. One of these constraints was derived from the population of the Jupiter-family 
comets (JFCs), as most recently described in Brasser \& Morbidelli (2013). The argument was used to estimate that there
were between $\sim$$2\times10^{11}$ and $\sim$$10^{12}$ objects in the original disk with $D>2.3$ km (Morbidelli \& Rickman 2015). 
If correct, it would require that the shallow size distribution below the break at $D\simeq100$ km needs to steepen up 
for small sizes. Here we satisfy this constraint by postulating a cumulative index of 3.0 for $1<D<10$ km. Note that this
contradicts the size distribution inferred from the observations of active JFCs, which is more shallow for 
$1<D<10$ km (cumulative index $\simeq$2; e.g., Lowry et al. 2008). At least part of this difference could presumably be explained by 
devolatization and surficial mass loss of cometary nuclei (Belton 2014). Finally, the detection of a single occultation event in 
the archival data of the HST guiding camera can be used to estimate the number of sub-kilometer KBOs (Schlichting et al. 2009). 
From this we infer that there would need to be $\sim$$10^{13}$-$10^{14}$ bodies with $D>0.5$ km in the original disk.  

The size distribution shown in Fig. \ref{sfd} was normalized to have $M_{\rm disk}=20$ $M_{\rm Earth}$, which is the preferred 
disk mass from NM12.  Different populations of small bodies in the Solar System have different probabilities to 
dynamically evolve from the original disk to reach their current orbits. For example, the capture probability of
Jupiter Trojans was estimated to be $P_{\rm JT} \simeq 7\times10^{-7}$ for each particle in the original disk (Nesvorn\'y 
et al. 2013). By scaling down by this factor the size distribution shown in Fig. \ref{sfd}, we find that the largest 
captured object should have $D \simeq 200$ km. For comparison, the largest Jupiter Trojan, 624 Hector, is roughly 
230 km accross. This illustrates that the normalization of the reconstructed profile from NM12 is consistent with the 
present population of Jupiter Trojans. Also, the probability that a disk planetesimal is captured as an irregular 
satellite of Jupiter is $\sim2\times10^{-8}$ according to Nesvorn\'y et al. (2014). This implies that the largest 
irregular satellite of Jupiter should have $D \sim 100$ km, while Himalia is only slightly larger 
($D \sim 140$ km).

The size distribution profile shown in Fig. \ref{sfd} has several interesting implications for the accretion and collisional 
evolution of KBOs. First, the hump in the profile at the largest sizes, with 1000-4000 Plutos, probably hints on a 
runaway-type mode of accretion of these largest objects. It is fairly similar to the size distribution profiles obtained 
in the classical collisional coagulation models (e.g., Stern \& Colwell 1997, Kenyon et al. 2008). It is unclear
whether the pebble accretion (e.g., Lambrechts \& Johansen 2012), which is a very efficient mechanism for growing large 
solid objects in the protoplanetary disks, could generate the hump. 

The size distribution at small sizes should have been modified by collisional grinding. The importance of collisional 
grinding mainly depends on the physical strenghts of KBOs, the dynamical structure of the outer planetesimal disk, and 
the time elapsed between the dispersal of the protoplanetary nebula and Neptune's migration into the disk. Using different 
assumptions, the published studies of collisional grinding reached different conclusions (e.g., Pan \& Sari 2004, 
Fraser 2009, Nesvorn\'y et al. 2011). If Neptune's migration into the planetesimal disk was delayed, 
as required if the planetary instability was responsible for the Late Heavy Bombardment (LHB; e.g., Gomes et al. 2005, 
Bottke et al. 2012), more time would be available for the modification of the size distribution by collisional grinding 
($\simeq$300-600 Myr, depending on when exactly the LHB started). Our main concern with this issue is whether a massive 
planetesimal disk could have survived a long period of collisional grinding, and have the estimated mass 
$M_{\rm disk}\simeq20$ $M_{\rm Earth}$ when the instability happened.

At least two important approximations were adopted in this work: (1) the gravitational effects of planetesimals  
were not explicitly included in the simulations (except for the implicit assumption that the small planetesimals drive
Neptune's migration and that the large planetesimals are the source of a jitter in the evolution of Neptune's
semimajor axis), and (2) the direct gravitational effects of the hypothetical fifth giant planet were 
not accounted for in the simulations except that we (optionally) activated Neptune's jump in some simulations to see 
whether Neptune's jump can resolve the resonance overpopulation problem. Here we argue that none of these assumptions can affect the 
main results of our work. As for (1), the collective gravitational effect of planetesimals can speed up the apsidal
and nodal precession of Neptune's orbits, and slightly alter the degree of the secular excitation of orbits in 
the Kuiper belt (Batygin et al. 2012). While this may be important to some extent for CCs, whose clustered orbital 
distribution would more easily reveal signs of small perturbations, this effect is probably insignificant for the 
resonant populations and HCs, which suffered much larger orbital changes due to other major dynamical processes.

As for (2), the five planet model of the early Solar System is, despite its various successes, not universally accepted 
and much more work will need to be done to establish things more firmly. It is thus probably sensible that here we did not
include the direct effects of the fifth planet on planetesimals. Instead, we showed in this work that the resonance overpopulation 
problem can be resolved if we include a reasonable number of Pluto-class planetesimals in the original trans-planetary
disk, and let Neptune's orbit react to the gravitational perturbations during close encounters with these bodies. About half 
of our simulations were done with Neptune's jump, which was presumably caused by an encounter of Neptune with the 
fifth planet (NM12). 

The basic motivation for activating Neptune's jump in some of our simulations was to test whether 
the jump can resolve, in itself, the resonance overpolulation problem as suggested in our previous work (Nesvorn\'y 2015a). We 
found that it cannot, because large populations of bodies are captured into resonances {\it after} Neptune's jump, during the 
subsequent migration of Neptune. Since the fifth planet was presumably ejected from the Solar System near the time of Neptune's
jump (NM12), it cannot affect things at later times. Including or ignoring its direct effects on planetesimals is therefore
irrelevant for the main thesis of this work. We plan on conducting more self-consistent simulations in the near future.
 
\acknowledgments
This work was supported by NASA's Outer Planet Research (OPR) program.
The work of David Vokrouhlick\'y was partly supported by the Czech Grant Agency (grant GA13-01308S). 
The CPU-expensive simulations in this work were performed on NASA's Pleiades Supercomputer, and on
the computer cluster Tiger at the Institute of Astronomy of the Charles University, Prague.

\clearpage
\begin{table}
\centering
{
\begin{tabular}{lrrr}
\hline \hline
                      & $P_{\rm HC}$                & $P_{3:2}$           & $N_{\rm HC}/N_{3:2}$  \\  
                      & ($\times10^{-4}$)    & ($\times10^{-4}$)         &         \\  
\hline
\multicolumn{4}{c}{\it Smooth Migration} \\
T30               & 1.9    & 5.3   & 0.36   \\     
C1-0.0            & 2.8    & 20    & 0.14   \\    
C1-0.5            & 2.4    & 6.8   & 0.35   \\             
C2-0.0            & 5.0    & 11    & 0.45   \\ 
C2-0.5            & 6.6    & 10    & 0.66   \\   

\hline
\multicolumn{4}{c}{\it Grainy Migration} \\
C1-0.0-1000P    & 4.6    & 8.9     &  0.52  \\
C1-0.5-1000P    & 4.2    & 3.2     &  1.3   \\  
C1-0.0-2000P    & 5.2    & 3.9     &  1.3   \\
C1-0.5-2000P    & 5.7    & 3.2     &  1.8    \\  
C1-0.0-4000P    & 6.6    & 2.1     &  3.1    \\
C1-0.5-4000P    & 6.2    & 1.9     &  3.3    \\  
C1-0.0-1000P2   & 5.5    & 1.3     &  4.2   \\
C1-0.5-1000P2   & 6.0    & 1.6     &  3.8    \\
C2-0.0-1000P    & 8.5    & 4.7     &  1.8      \\
C2-0.5-1000P    & 9.2    & 3.7     &  2.5     \\  
C2-0.0-2000P    & 11     & 1.9     &  5.8  \\
C2-0.5-2000P    & 13     & 1.9     &  6.8  \\  

\hline
\multicolumn{4}{c}{\it Observations} \\
CFEPS/DES       & $\simeq$5 & $\simeq$1.5     &  $\simeq$2-4  \\  
                        
\hline \hline
\end{tabular}
}
\caption{The capture statistics of the HCs and Plutinos obtained in different dynamical models. T30 is a case from Nesvorn\'y (2015a),
where Neptune started at $a_{\rm N,0}=24$ AU and smoothly migrated to 30 AU with an e-folding timescale of $\tau=30$ Myr. C1 stands for case 1 with 
$\tau_1=30$~Myr and $\tau_2=100$ Myr, C2 stands for case 2 with $\tau_1=10$ Myr and $\tau_2=30$~Myr. Labels 0.0 and 0.5 denote the cases with 
$\Delta a_{\rm N}=0.0$ and  $\Delta a_{\rm N}=0.5$ AU. The simulations with 1000, 2000 and 4000 Plutos are labeled by 1000P, 2000P and 4000P,
respectively. The case with 1000 Twoplutos is denoted by 1000P2. The columns give the probability of capture
as HC ($P_{\rm HC}$) and in the 3:2 resonance ($P_{3:2}$), and the ratio between the two populations ($N_{\rm HC}/N_{3:2}$).
The last row lists observational contraints. The $P_{\rm HC}$ value reported in this row was computed from the estimated mass of the 
HCs, $\sim$0.01 $M_{\rm Earth}$ according to Fraser et al. (2014). Assuming a $M_{\rm disk}=20$ $M_{\rm Earth}$ disk from NM12, this 
gives $P_{\rm HC} \simeq 5 \times 10^{-4}$. According to the CFEPS and DES the population of Plutinos in the 3:2 resonance is
$\simeq$2-4 smaller than the HCs. Thus, $P_{\rm HC} \simeq 1.5 \times 10^{-4}$.}
\end{table}

\clearpage

\clearpage

\clearpage

\clearpage

\clearpage



\clearpage

\clearpage

\clearpage


\clearpage

\clearpage

\clearpage
\begin{figure}
\epsscale{0.45}
\plotone{fig11a.eps}
\plotone{fig11b.eps}
\caption{A comparison of the inclination distributions obtained in our model (solid lines) and the CFEPS 
detections (dashed lines). Here we used $\tau_1=10$ Myr, $\tau_2=30$ Myr, $a_{\rm N,1}=27.8$ AU, $\Delta 
a_{\rm N} =0.5$ AU, $\Delta e_{\rm N} = 0.1$, and the migration graininess corresponding to 2000 massive
planetesimals each with $M_{\rm mp}=M_{\rm Pluto}$. The CFEPS detection simulator was applied to the model 
orbits to have a one-to-one comparison with the actual CFEPS detections. In panel (b), we plot orbits with 
$i>5^\circ$.}
\label{idist2}
\end{figure}

\clearpage
\begin{figure}
\epsscale{0.8}
\plotone{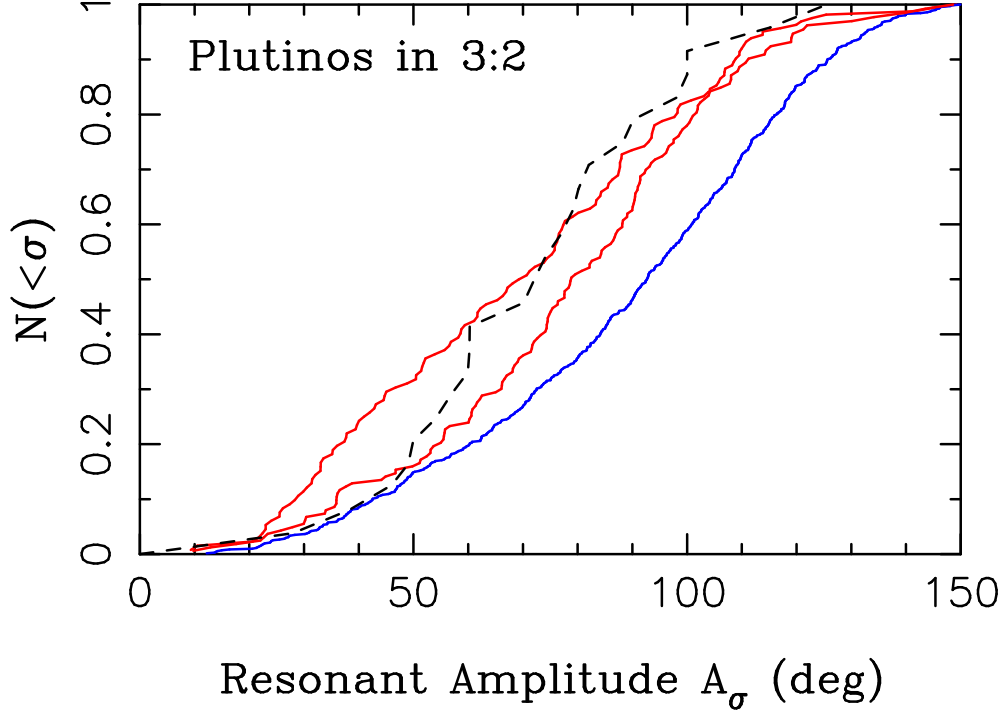}
\caption{The cumulative distributions of the libration amplitudes in the 3:2 resonance. The dashed line shows 
the distribution from CFEPS (Gladman et al. 2012). The blue solid line is the distribution obtained in the smooth
migration case with $\tau_1=30$ Myr, $\tau_2=100$ Myr, and $a_{\rm N,1}=27.8$ AU. The two red lines show the 
distributions for the same model parameters, but when we assumed that the planetesimal disk contained 1000
massive planetesimals each with $M_{\rm mp}=2\ M_{\rm Pluto}$. The two cases correspond to $\Delta a =0$ (shallower 
profile) and $\Delta a =0.5$ AU (steeper profile).}
\label{sigma}
\end{figure}

\clearpage
\begin{figure}
\epsscale{0.8}
\plotone{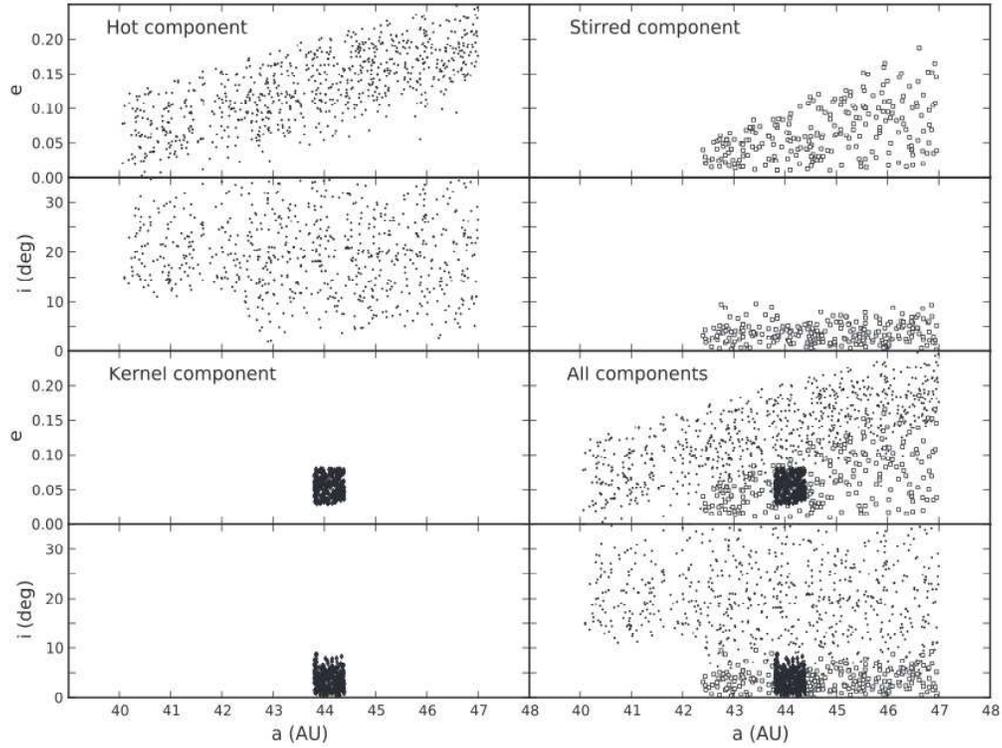}
\caption{Three components of the CFEPS-L7 synthetic model for the main classical belt. The hole at $a\simeq40$-42 AU and 
low inclinations in the hot component was introduced to represent the destabilizing action of the secular resonances.
The kernel component is the concentration of orbits with $43.8<a<44.4$ AU, $0.03<e<0.08$ and $i\lesssim5^\circ$. 
Figure from Petit et al. (2011).}
\label{kernel}
\end{figure}

\clearpage
\begin{figure}
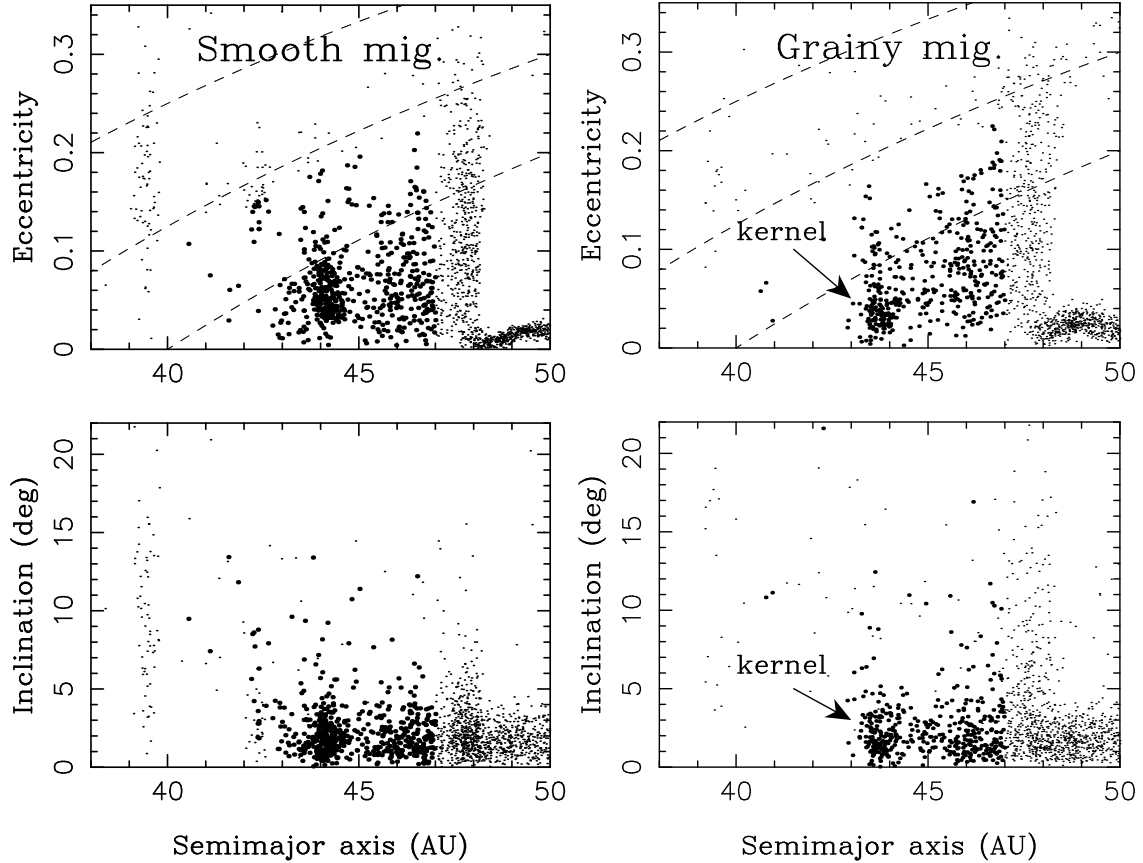

\epsscale{0.45}
\plotone{fig14a.eps}
\plotone{fig14b.eps}
\caption{The final distribution of orbits obtained in two simulations with $a_{\rm N,0}=24$ AU, $\tau_1=30$ Myr, 
$a_{\rm N,1}=27.8$ AU, $\Delta a_{\rm N} = 0.5$ AU, $\Delta e_{\rm N} = 0.1$, and $\tau_2=100$ Myr. The panels on the 
left show the result for the smooth migration (figure from Nesvorn\'y 2015b), while those on the right show the 
result for the grainy migration with 1000 massive planetesimals each with $M_{\rm mp}=2\ M_{\rm Pluto}$.
The concentration of orbits at $\simeq$44 AU was created by the 2:1 resonance when
Neptune jumped. At the beginning of the simulation, 5000 test particles were distributed on low-inclination 
($\sigma_i=2^\circ$) low-eccentricity ($\sigma_e=0.01$) orbits between 30 and 50 AU. The bold symbols denote the 
orbits that ended with $40<a<47$ AU and $q=a(1-e)>36$ AU.}
\label{colds}
\end{figure}

\clearpage
\begin{figure}
\epsscale{0.7}
\plotone{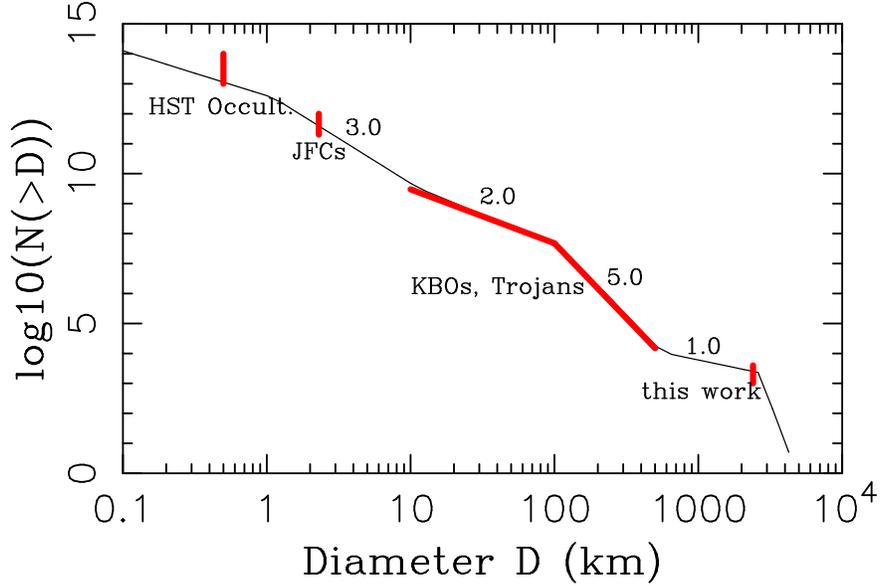}
\caption{A schematic reconstruction of the size distribution of the original planetesimal disk below 30 AU. The red 
color denotes various constraints. HST Occult. stands for the occultation constraint derived in Schlichting et al.
(2009), JFCs is the constraint from Morbidelli \& Rickman (2015), and the distribution for $10<D<500$ km
is inferred from the observations of KBOs and Jupiter Trojans (e.g., Fraser et al. 2014). The break between a shallow 
slope for small sizes and a steep slope for large sizes was fixed at $D=100$~km. The existence of 1000-4000 
Plutos in the original disk inferred in this work requires that the size distribution had a hump at $D>500$ km. 
The numbers above the reconstructed size distribution show the cumulative power index that was used for different 
segments. The wavy nature of the size distribution shown here is reminiscent of that of the present asteroid belt.}
\label{sfd}
\end{figure}

\end{document}